\documentclass[sigconf]{acmart}

\usepackage[T1]{fontenc}
\usepackage[utf8]{inputenc}
\usepackage{tabularx}        % 너비 조절 가능한 테이블
\usepackage{makecell}        % 테이블 셀 안에서 줄바꿈
\usepackage{float}           % [H] 옵션으로 그림 위치 고정
\usepackage{wrapfig}         % 텍스트로 그림/표 감싸기
\usepackage{ragged2e}        % \RaggedRight 등 향상된 정렬
\usepackage{cuted}      % <-- 이 줄을 추가하세요.

\usepackage{graphicx}        % acmart가 이미 로드하지만, 명시적으로 포함
\usepackage[graphicx]{realboxes} % realboxes 사용을 위해 graphicx 옵션 추가

\usepackage{pifont} 
\usepackage{enumitem}        % 리스트 환경 커스터마이징
\usepackage[most]{tcolorbox} % 고급 색상 박스
\definecolor{titlefg}{HTML}{FFFFFF}
\definecolor{titlebg}{HTML}{0D47A1}
\definecolor{framec}{HTML}{B0BEC5}
\definecolor{backc}{HTML}{F7F9FB}
\definecolor{accent}{HTML}{1565C0}
\definecolor{impact}{HTML}{F6A623}
\definecolor{buyc}{HTML}{2E7D32}
\definecolor{sellc}{HTML}{C62828}
\definecolor{momentumc}{HTML}{1565C0}
\definecolor{contrac}{HTML}{F57C00}

\definecolor{TopStrong}{HTML}{B2DFDB}
\definecolor{BottomStrong}{HTML}{F9D5D3}
\definecolor{TextBlue}{HTML}{B2DFDB}
\definecolor{TextRed}{HTML}{F9D5D3}

\newcommand{\IMPACT}{\textbf{\textcolor{impact}{5\%}}}
\newcommand{\HRule}{\noindent\textcolor{accent!40}{\rule{\linewidth}{0.6pt}}}
\newcommand{\hlinline}[1]{\colorbox{yellow!30}{\strut #1}}

\newtcbox{\numbadge}{on line,
  colback=accent!12, coltext=accent!85, boxrule=0pt,
  left=0.45ex, right=0.45ex, top=0.05ex, bottom=0.05ex,
  arc=0.8ex, tcbox raise base}
\newcommand{\nb}[1]{\numbadge{\bfseries\scriptsize #1}}

\newtcbox{\tagpill}[1][]{on line, boxrule=0pt, arc=1ex,
  top=0.1ex, bottom=0.1ex, left=0.7ex, right=0.7ex,
  tcbox raise base, colback=black!6, coltext=black!80, #1}
\newcommand{\Momentum}{\tagpill[colback=momentumc!15,coltext=momentumc!90]{\bfseries Momentum}}
\newcommand{\Contrarian}{\tagpill[colback=contrac!15,coltext=contrac!90]{\bfseries Contrarian}}
\newtcbox{\keypill}{on line, boxrule=0pt, arc=0.9ex,
  top=0.1ex, bottom=0.1ex, left=0.55ex, right=0.55ex,
  tcbox raise base, colback=black!10, coltext=black!65}

\newtcolorbox{promptblock}[2][]{enhanced, breakable,
  colback=backc, colframe=framec, boxrule=0.8pt, arc=2mm,
  title={#2}, colbacktitle=titlebg, coltitle=titlefg,
  fonttitle=\bfseries\ttfamily, title filled, titlerule=0pt,
  top=3.5mm, bottom=4mm, left=5mm, right=5mm,
  before upper={\RaggedRight\setlength{\emergencystretch}{2em}},
  #1}

\newtcolorbox{evsection}[2][]{%
  enhanced, breakable,
  colback=white, colframe=#2!28,
  borderline west={2pt}{0pt}{#2!65},
  boxrule=0.5pt, arc=1mm,
  left=2mm, right=2mm, top=0.6mm, bottom=0.6mm,
  #1
}
\newtcbox{\chip}[1][]{on line, boxrule=0pt, arc=1ex,
  top=0.1ex, bottom=0.1ex, left=0.8ex, right=0.8ex,
  tcbox raise base, colback=black!6, coltext=black!75, #1}  

\copyrightyear{2025}
\acmYear{2025}
\setcopyright{cc}
\setcctype{by}
\acmConference[ICAIF '25]{6th ACM International Conference on AI in Finance}{November 15--18, 2025}{Singapore, Singapore}
\acmBooktitle{6th ACM International Conference on AI in Finance (ICAIF '25), November 15--18, 2025, Singapore, Singapore}\acmDOI{10.1145/3768292.3770375}
\acmISBN{979-8-4007-2220-2/2025/11}

\begin{document}
             
\title[Your AI, Not Your View: The Bias of LLMs in Investment Analysis]{{Your AI, Not Your View: The Bias of LLMs in Investment Analysis}}

\author{Hoyoung Lee}
\affiliation{
  % \institution{Ulsan National Institute of Science and Technology}
  \institution{UNIST}
  \city{Ulsan}
  \country{Republic of Korea}}
\email{hoyounglee@unist.ac.kr}

\author{Junhyuk Seo}
\affiliation{
  \institution{UNIST}
  \city{Ulsan}
  \country{Republic of Korea}}
\email{brians327@unist.ac.kr}

\author{Suhwan Park}
\affiliation{
  \institution{UNIST}
  \city{Ulsan}
  \country{Republic of Korea}}
\email{suhwan@unist.ac.kr}

\author{Junhyeong Lee}
\affiliation{
  \institution{UNIST}
  \city{Ulsan}
  \country{Republic of Korea}}
\email{jun.lee@unist.ac.kr}

\author{Wonbin Ahn}
\affiliation{
  \institution{LG AI Research}
  \city{Seoul}
  \country{Republic of Korea}
}
\email{wonbin.ahn@lgresearch.ai}

\author{Chanyeol Choi}
\affiliation{
  \institution{LinqAlpha}
  \city{New York}
  \country{United States}}
\email{jacobchoi@linqalpha.com}

\author{Alejandro Lopez-Lira}
\affiliation{
  \institution{University of Florida}
  \city{Gainesville}
  \country{United States}}
\email{alejandro.lopez-lira@warrington.ufl.edu}

\author{Yongjae Lee}
\authornote{Corresponding author.}
\affiliation{
  \institution{UNIST}
  \city{Ulsan}
  \country{Republic of Korea}}
\email{yongjaelee@unist.ac.kr}

\renewcommand{\shortauthors}{Lee et~al.}

% ------------------------------------------------------------
% \linenumbers

% — Abstract —
\begin{abstract}
{In finance, Large Language Models (LLMs) face frequent knowledge conflicts arising from discrepancies between their pre-trained parametric knowledge and real-time market data. These conflicts are especially problematic in real-world investment services, where a model's inherent biases can misalign with institutional objectives, leading to unreliable recommendations. Despite this risk, the intrinsic investment biases of LLMs remain underexplored. We propose an experimental framework to investigate emergent behaviors in such conflict scenarios, offering a quantitative analysis of bias in LLM-based investment analysis. Using hypothetical scenarios with balanced and imbalanced arguments, we extract the latent biases of models and measure their persistence. Our analysis, centered on sector, size, and momentum, reveals distinct, model-specific biases. Across most models, a tendency to prefer technology stocks, large-cap stocks, and contrarian strategies is observed. These foundational biases often escalate into confirmation bias, causing models to cling to initial judgments even when faced with increasing counter-evidence. A public leaderboard benchmarking bias across a broader set of models is available at \url{https://linqalpha.com/leaderboard}.}
\end{abstract}

\begin{CCSXML}
<ccs2012>
   <concept>
       <concept_id>10010147.10010178.10010179</concept_id>
       <concept_desc>Computing methodologies~Natural language processing</concept_desc>
       <concept_significance>500</concept_significance>
       </concept>
   <concept>
       <concept_id>10010405.10010455.10010460</concept_id>
       <concept_desc>Applied computing~Economics</concept_desc>
       <concept_significance>500</concept_significance>
       </concept>
 </ccs2012>
\end{CCSXML}

\ccsdesc[500]{Computing methodologies~Natural language processing}
\ccsdesc[500]{Applied computing~Economics}

\keywords{Large Language Models, Financial Bias, Knowledge Conflict, Trustworthy AI, Decision-Making, Investment Analysis, Preference}

\maketitle

% \footnotetext[1]{https://linqalpha.com/leaderboard}
\section{Introduction}
The rapid advancement of LLMs has spurred a surge of innovation within the financial sector, where they are particularly adept at processing qualitative and unstructured information. Research is now actively exploring their use across a range of applications, including forecasting stock price movements from news sentiment \cite{lo2024can}, extracting nuanced insights from complex analyst reports \cite{kim2023llms}, and assisting in the construction and optimization of portfolios \cite{hwang2025decision, lee2025integrating, ko2024can}. This trend is now evolving towards even greater autonomy through the development of sophisticated LLM-based agents. These systems, which may function as a single powerful agent or as collaborative multi-agent teams, are designed to execute complex, dynamic tasks like active trading and automated portfolio management \cite{zhang2024multimodal, yu2024fincon}.

\begin{figure}[t]
\centering
\includegraphics[width=\columnwidth]{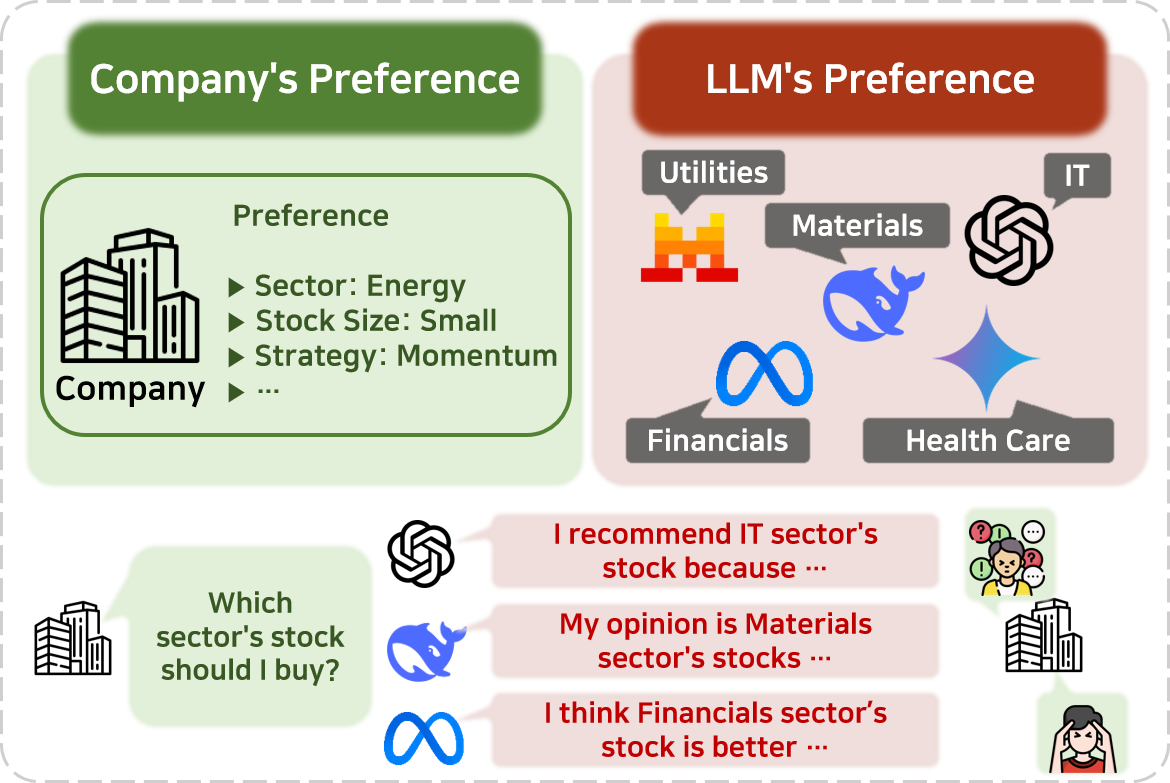} 
\caption{A conceptual illustration of knowledge conflict in LLM-based financial services. Even when a firm targets a specific investment theme (e.g., Energy), the LLM’s inherent preferences (e.g., Technology) may override user intent, producing biased and inconsistent recommendations.}
\label{fig:llm_preference_conflict}
\end{figure}

A critical but underexplored issue in financial applications is \textbf{knowledge conflict}. In a domain as fluid and time-sensitive as finance, conflicts between an LLM's parametric knowledge and real-time market data are frequent. This conflict becomes particularly revealing when the model is presented with a mix of information. Crucially, studies show that when an LLM encounters both supporting evidence (aligning with its ingrained beliefs) and counter-evidence simultaneously, it exhibits a strong confirmation bias \cite{xie2023adaptive}. Instead of weighing the arguments objectively, the model stubbornly adheres to the evidence that confirms its internal knowledge while disregarding the counter-evidence.

This tendency to reinforce internal biases over objective reasoning poses a major risk to LLM-based financial services. For instance, as illustrated in Figure~\ref{fig:llm_preference_conflict}, even if a financial institution wants to target a specific sector (e.g., Energy), the LLM may override this with its own preference (e.g., Technology). Consequently, this creates a dilemma: the service reflects the model’s bias, not the user’s intent, leading to distorted, unpredictable decisions that ultimately erode client trust.

To address this core problem, we must first systematically uncover these hidden biases. We therefore seek to answer the following research questions:

\begin{itemize}
\item[\textbf{RQ 1:}] What biases do LLMs exhibit towards key financial factors like sector, size, and momentum?
\item[\textbf{RQ 2:}] What issues arise from these biases when LLMs are forced to make decisions with conflicting evidence?
\end{itemize}

To answer these questions, this study introduces a three-stage experimental framework designed to systematically elicit and verify LLM biases in investment analysis. In the first stage, we generate equally compelling but opposing arguments for each stock, such as positive vs. negative sentiment or momentum vs. contrarian perspectives, to represent competing investment views. In the second stage, we present these arguments in a balanced manner to induce a knowledge conflict and reveal the model’s underlying biases. In the third stage, we introduce progressively stronger counter-evidence to examine the problems that arise from these biases, observing how they evolve into rigid, confirmation-biased judgments.

The main contributions of this paper are twofold. First, we propose a systematic methodology to identify and quantify latent biases in LLMs for financial applications. Second, we provide the quantitative analysis of confirmation bias exhibited by LLMs in investment analysis, demonstrating a clear link between a model's inherent biases and the critical issues that arise, such as a stubborn refusal to revise judgments despite increasing counter-evidence. This work serves as a critical reminder that for financial AI, trustworthiness is a benchmark as vital as performance. By systematically uncovering these hidden risks, our work lays a critical foundation for developing more transparent and trustworthy financial AI. 
\section{Background}
This section provides background on knowledge conflict and financial biases in LLMs to contextualize our investigation into their intersection.

\subsection{Knowledge Conflict}
A critical vulnerability in LLMs is knowledge conflict, which arises when external, contextual information clashes with the model's internal, parametric knowledge \cite{sun2025seen, jin2024tug}. A significant body of research demonstrates that when faced with such conflicts, LLMs exhibit a strong confirmation bias. Foundational work by \cite{xie2023adaptive} revealed that LLMs behave as "stubborn sloths," clinging to any piece of evidence that supports their internal knowledge, even against a majority of contradictory facts. This over-reliance on internal memory is further evidenced by findings that LLMs struggle to suppress their parametric knowledge even when instructed to \cite{sun2025seen} and can exhibit a Dunning-Kruger-like effect, confidently trusting their own faulty beliefs over correct external information \cite{jin2024tug}.

This tendency toward knowledge-based stubbornness is part of a broader pattern. Given their training on vast amounts of human data, LLMs have been shown to inherit and functionally replicate human cognitive biases \cite{echterhoff2024cognitive}. A key example is the choice-supportive bias, where the mere act of making an initial choice significantly boosts the model's confidence in that choice, making it highly resistant to change \cite{kumaran2025overconfidence, zhuang2025llm}. This phenomenon is part of a wider landscape of biases identified in LLMs when they act as evaluators. For instance, models exhibit familiarity bias (preferring text they find easier to process), are susceptible to anchoring effects \cite{stureborg2024large}, and can be biased towards their own generated contexts over externally retrieved information, even when their own generated text is incorrect \cite{tan2024blinded}.

\subsection{Financial Biases in LLMs}
The presence of these biases is particularly concerning in the economic and financial domains \cite{bini2025behavioral, cao2025llms, chen2024what, cook2025social}. Recent evidence suggests that alignment tuning can shift LLMs’ risk preferences, influencing their financial decision-making \cite{ouyang2024risk}. Initial research has begun to map their characteristics, with frameworks applying utility theory demonstrating that LLMs are neither perfectly rational nor consistently human-like \cite{ross2024llm}. Other studies note that even specialized financial LLMs can exhibit strong irrationalities \cite{zhou2024llms}. While this foundational work is critical for establishing the existence of such biases, the methodologies employed often diverge from the complex process of real-world investment analysis. For instance, biases have been identified by measuring how a firm’s name alters sentiment in a single sentence \cite{nakagawa2024evaluating} or by detecting preferential stock recommendations across numerous scenarios \cite{zhi2025exposing}.

However, existing simplified approaches overlook the core of real-world investment analysis: synthesizing conflicting signals. We address this gap by designing a realistic testbed that presents LLMs with balanced, contradictory arguments. By progressively introducing stronger counter-evidence, we then observe how initial biases lead to flawed decision-making. This approach provides a systematic framework for diagnosing the practical risks posed by LLM biases in finance.
\section{Methodology}
\begin{figure}[htbp]
    \centering
    \includegraphics[width=\columnwidth]{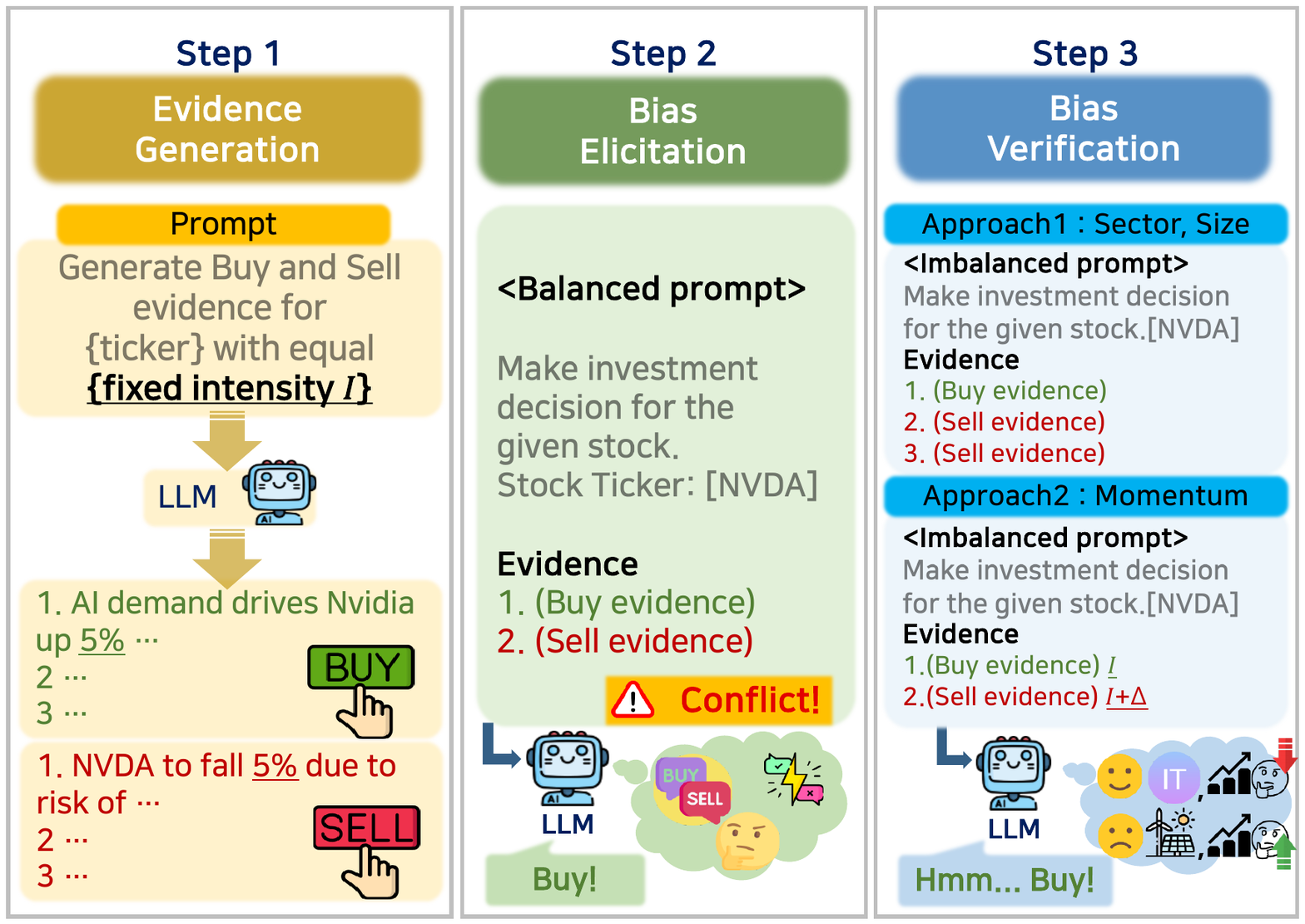}
    \caption{The three-stage experimental framework: (1) Generating balanced evidence, (2) Eliciting bias through knowledge conflict, and (3) Verifying the resulting bias against counter evidence.}
    \label{fig:figure2}
\end{figure}

This study adopts a three-stage experimental framework, as illustrated in Figure~\ref{fig:figure2}, to examine whether biases in LLMs cause skewed financial decisions. All experiments utilize a standardized prompt structure, $\mathcal{P} = (T, C, A)$, comprising three components: a fixed Task ($T$) instructing the model to make an investment decision, a variable Context ($C$) containing the evidence set, and a fixed set of permissible Actions ($A$) defined as $\{\text{buy}, \text{sell}\}$. Our methodology is designed to probe the model's behavior when faced with conflicting information within this framework.

\subsection{Experimental Setup}
To isolate and analyze biases rooted in the model's parametric knowledge, our experimental design aims to mitigate the risk of hallucination. This approach is based on evidence suggesting that models are significantly less prone to generating fabricated information when prompted about subjects they are familiar with from their training data \cite{ferrando2024know}.

Accordingly, our investigation is confined to a curated set of 427 prominent stocks, denoted $\mathcal{S} = \{s_1, s_2, \dots, s_{427}\}$. These stocks were selected for their continuous listing in the S\&P 500 index over the past five years. Their high public visibility increases the likelihood that they are well-represented in the models' training corpora, thus grounding the experiment in internal knowledge rather than speculative generation. All experiments were performed with the models configured at a temperature of $\tau = 0.6$, striking a balance between deterministic and creative response generation, and each experimental run was conducted three times to ensure the robustness of our results. Detailed specifications of the models used are provided in Appendix~\ref{sec:appendix_models}.

\begin{figure*}[!t]
  \centering
  \includegraphics[width=\textwidth]{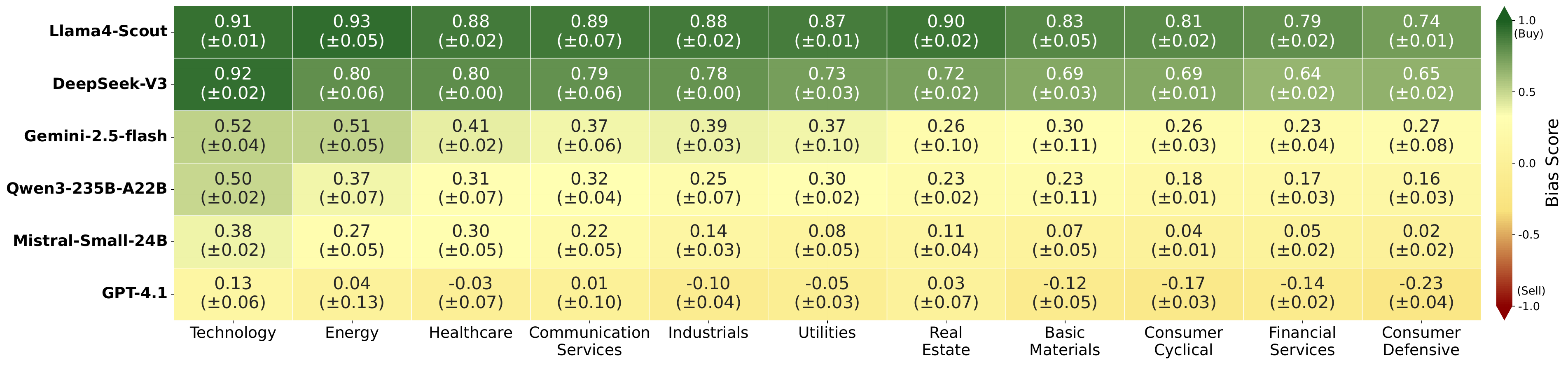}
\caption{Sector bias scores for each evaluated LLM. Scores represent the mean of three independent sets of 10 trials; the standard deviation in parentheses reflects the variation across the three sets. Green indicates a positive (buy) bias and red a negative (sell) bias. A strong bias toward the Technology sector is evident in most models.}
  \label{fig:sector-heatmap}
\end{figure*}

\subsection{Evidence Generation}
To construct balanced qualitative and quantitative arguments for each stock $s \in \mathcal{S}$, we leverage \texttt{Gemini-2.5-Pro} \cite{comanici2025gemini}, a model that is deliberately chosen to be separate from the six LLMs under evaluation. This design ensures neutrality in evidence generation, minimizing alignment with any of the test models. Recent work has highlighted that LLMs can exhibit a strong bias towards LLM-generated content over externally retrieved information \cite{tan2024blinded}. To neutralize this potential generation bias and ensure that observed bias are not artifacts of context sourcing, our methodology exclusively uses generated evidence for all experimental conditions.

Specifically, for every stock $s$, buy evidence ($\mathcal{E}_{\text{buy}}^{(s)}$) and sell evidence ($\mathcal{E}_{\text{sell}}^{(s)}$) are generated in an equal proportion, yielding a comprehensive dataset of $|\mathcal{E}| = 3,416$ evidence. To further isolate bias, all evidence is engineered with a uniform linguistic structure and explicitly states a fixed intensity parameter $I = 5\%$. Thus, each buy evidence posits an expected price appreciation of $I$, while each sell evidence anticipates a depreciation of $I$:
\begin{equation}
\label{eq:evidence_intensity_cases}
\mathbb{E}[\Delta p^{(s)}] =
\begin{cases}
    +I & \text{for a buy evidence } e_{\text{buy},i}^{(s)} \\
    -I & \text{for a sell evidence } e_{\text{sell},i}^{(s)}
\end{cases}
\end{equation}
where $\Delta p^{(s)}$ represents the projected price change for stock $s$.

\subsection{Bias Elicitation}
\label{sec:preference_elicitation}
This stage aims to elicit the LLM's underlying bias by leveraging the confirmation bias that emerges during knowledge conflicts. When an LLM is presented with conflicting information, it may exhibit a tendency to favor evidence that aligns with its parametric knowledge. We deliberately engineer such a conflict using a \textbf{balanced prompt}. The context $C_s$ of this prompt contains an equal proportion of buy and sell evidence ($|\mathcal{E}_{\text{buy}}^{(s)}| = |\mathcal{E}_{\text{sell}}^{(s)}|$), each with the same intensity. In this state of informational equilibrium, where external evidence is mutually contradictory, the model's ultimate decision is hypothesized to be guided by its internal parametric memory regarding the stock $s$. The resulting choice thereby reveals its inherent bias. See Appendix~\ref{sec:prompt_example} for detailed prompt example.

To quantify this elicited bias, the decision task is repeated $N = 10$ times for each stock, with the evidence order randomized in each trial to mitigate positional bias. This yields decision counts $N_{\text{buy}}^{(s)}$ and $N_{\text{sell}}^{(s)}$, from which the bias score($\pi_s$) is calculated as:
\begin{equation}
\label{eq:preference_score}
\pi_s = \frac{N_{\text{buy}}^{(s)} - N_{\text{sell}}^{(s)}}{N_{\text{buy}}^{(s)} + N_{\text{sell}}^{(s)}},
\end{equation}
where the resulting score $\pi_s$ ranges from -1 to 1. A value approaching 1 indicates a pronounced bias towards buying, while a value approaching -1 indicates a pronounced bias towards selling.

\subsection{Bias Verification}
\label{sec:bias_verification}

The goal of this stage is to verify the problems that arise from groups exhibiting high bias scores. First, we partition the set of all stocks $\mathcal{S}$ into disjoint groups (e.g., by market sector) and identify the group $\mathcal{G}^*$ that exhibits the highest average bias score.

To test what problems this group-level bias causes in the decision-making process, let $s^*$ denote any stock from this most-biased group ($s^* \in \mathcal{G}^*$). Evidence that aligns with the group's established bias for $s^*$ (e.g., buy evidence for a buy-biased group) is termed supporting evidence. Conversely, evidence that opposes this bias is designated as counter-evidence. We then subject each stock $s^* \in \mathcal{G}^*$ to a test using an \textbf{imbalanced prompt}. This is a prompt where the counter-evidence is deliberately strengthened—either in volume or intensity—to challenge the model's initial bias. We then measure the decision flip rate, $\phi_{s^*}$. This verification is conducted from two perspectives: evidence volume and evidence intensity.

\subsubsection{Approach 1: Verification by Evidence Volume}
One approach to assess bias tenacity is by creating a volumetric imbalance, presenting more counter-evidence than supporting evidence. For example, in a test case for a stock $s^* \in \mathcal{G}^*$ from a buy-biased group, the imbalanced context might contain two pieces of supporting evidence ($|\mathcal{E}_{\text{buy}}^{(s^*)}| = 2$) and three pieces of counter evidence ($|\mathcal{E}_{\text{sell}}^{(s^*)}| = 3$). The flip rate is computed as:
\begin{equation}
\label{eq:flip_rate_vol}
\phi_{s^*}^{\text{vol}} = \frac{N_{\text{flip}}^{(s^*)}}{N},
\end{equation}
where $N_{\text{flip}}^{(s^*)}$ counts instances where the original preference is overturned by the volumetric majority of counter evidence. A low $\phi_{s^*}^{\text{vol}}$ signifies that the model's intrinsic bias is strong enough to outweigh the volumetric majority of counter-evidence.

\subsubsection{Approach 2: Verification by Evidence Intensity}
An alternative approach is to test the model against counter-evidence of a fixed higher intensity while maintaining volumetric parity. For a stock $s^* \in \mathcal{G}^*$ from a buy-biased group, supporting evidence is presented at a standard baseline intensity, $I$, while counter-evidence is presented at an intensified level of $I + \Delta$.

This creates asymmetric conflict. For example, for a baseline intensity of $I = 5\%$ and an increment of $\Delta = 5\%$, the intensified level would be $10\%$. The expectations are:
\begin{equation}
\label{eq:intensity_conflict}
\begin{aligned}
\text{(Supporting Evidence)} \quad & e_{\text{buy},i}^{(s^*)}: \mathbb{E}[\Delta p^{(s^*)}] = +I, \\
\text{(Counter-Evidence)} \quad & e_{\text{sell},i}^{(s^*)}: \mathbb{E}[\Delta p^{(s^*)}] = -(I + \Delta).
\end{aligned}
\end{equation}
The intensity-driven flip rate is then measured as:
\begin{equation}
\label{eq:flip_rate_int}
\phi_{s^*}^{\text{int}} = \frac{N_{\text{flip}}^{(s^*)}}{N}.
\end{equation}
A low $\phi_{s^*}^{\text{int}}$ implies that the model's intrinsic bias is sufficiently tenacious to override even qualitatively stronger counter-evidence.

\section{Results}
% This section presents our empirical findings in sequence with our research questions. First, Section \ref{subsec:intrinsic_prefs} addresses RQ1 by identifying the intrinsic preferences of LLMs for stock attributes (sector, size) and investment styles (momentum). To validate these observed differences, we conducted statistical tests to quantify the significance of these preferences. Next, Section \ref{subsec:bias} addresses RQ2 by testing if these preferences become systematic biases under contradictory evidence, using \textit{Approach 1} for attributes and \textit{Approach 2} for style. Finally, Section \ref{subsec:uncertainty} analyzes the link between preference strength and the model's internal uncertainty, as measured by entropy.
This section presents our empirical findings, structured to answer our research questions sequentially. First, to address RQ1, Section \ref{subsec:bias_elicitation} identifies and quantifies the LLM's latent biases toward factors like sector, size, and momentum, confirming their statistical significance. Next, Section \ref{subsec:bias_verification} addresses RQ2 by quantitatively analyzing the resulting bias. We demonstrate how these biases lead to a stubborn refusal to revise judgments when the model is challenged with stronger counter-evidence in terms of both volume and intensity. Finally, Section \ref{subsec:uncertainty} further investigates this phenomenon by linking the strength of the bias to the model's internal uncertainty, as measured by entropy.

\subsection{Intrinsic Bias of LLMs}
\label{subsec:bias_elicitation}
\subsubsection{Sector Bias}
The bias scores presented in this section were calculated over 10 trials for each stock. To ensure the robustness of our findings against the inherent variability of LLM responses, we conducted this entire process in three independent sets; all results are reported as a mean score with the corresponding standard deviation.
Our analysis of inherent sector bias reveals two notable patterns: a prevalent bias toward the Technology sector and an overall tendency for positive bias scores, reflecting a default inclination to buy. However, the strength of this bias varies considerably across models (Figure~\ref{fig:sector-heatmap}). For example, models such as \texttt{Llama4-Scout} and \texttt{DeepSeek-V3} follow the general trend, consistently exhibiting high positive bias scores across most sectors. In contrast, other models display more nuanced behavior; \texttt{GPT-4.1}, for instance, not only shows a lower average bias but also exhibits negative bias scores in certain sectors, deviating from the common buy-oriented tendency.

To quantify these differences, we conducted independent samples t-tests comparing the mean bias scores between each model's highest and lowest bias sectors (Table~\ref{tab:ttest_sector}). Crucially, the analysis confirms that the gap between the most and least favored sectors is statistically significant for every model evaluated. This universal finding provides strong evidence that all models possess distinct and deeply embedded sector biases in their knowledge representations.

% Ultimately, our findings indicate that while a general 'buy' tendency is common, the strength and even the direction of sector biases are highly model-dependent. This underscores the critical importance of auditing and selecting LLMs for their inherent biases, especially for deployment in sensitive, real-world applications like financial analysis where model objectivity is paramount.

\begin{table}[htbp]
\caption{Independent-samples t-test results for the score gap between each model’s highest- and lowest-scoring sectors.}
\label{tab:ttest_sector}
\centering
\resizebox{\columnwidth}{!}{%
\begin{tabular}{lccc}
\specialrule{1.1pt}{0pt}{0pt}
\textbf{Model} & \textbf{High-Score} & \textbf{Low-Score} & \textbf{p-value} \\
\specialrule{1pt}{0pt}{0pt}
\texttt{Llama4-Scout} & Energy & Consumer Defensive & $<0.001^{***}$ \\
\texttt{DeepSeek-V3} & Technology & Financial Services & $<0.001^{***}$ \\
\texttt{Gemini-2.5-flash} & Technology & Financial Services & $<0.001^{***}$ \\
\texttt{Qwen3-235B} & Technology & Consumer Defensive & $<0.001^{***}$ \\
\texttt{Mistral-Small} & Technology & Consumer Defensive & $<0.001^{***}$ \\
\texttt{GPT-4.1} & Technology & Consumer Defensive & $<0.001^{***}$ \\
\specialrule{1.1pt}{0pt}{0pt}
\end{tabular}%
}
\flushleft{\footnotesize $^{*}p < 0.05$, $^{**}p < 0.01$, $^{***}p < 0.001$}
\end{table}

% \begin{table}[htbp]
% \caption{Independent samples t-test of the preference gap between the highest and lowest preference sectors. The Diff column shows the magnitude of the preference gap.}
% \label{tab:ttest_sector}
% \centering
% \resizebox{\columnwidth}{!}{%
% \begin{tabular}{lcccc}
% \specialrule{1.1pt}{0pt}{0pt}
% \textbf{Model} & \textbf{High-Pref} & \textbf{Low-Pref} & \textbf{Diff} & \textbf{p-value} \\
% \specialrule{1pt}{0pt}{0pt}
% \texttt{Llama4-Scout} & Energy & Consumer Defensive & 0.2064 & $<0.001^{***}$ \\
% \texttt{DeepSeek-V3} & Technology & Basic Materials & 0.2090 & 0.014$^{*}$ \\
% \texttt{Qwen3-235B} & Utilities & Consumer Cyclical & 0.2361 & 0.003$^{**}$ \\
% \texttt{Gemini-2.5} & Energy & Basic Materials & 0.2035 & 0.035$^{*}$ \\
% \texttt{GPT-4.1} & Energy & Communication Services & 0.1398 & 0.091 \\
% \texttt{Mistral-24B} & Basic Materials & Communication Services & 0.1444 & 0.124 \\
% \specialrule{1.1pt}{0pt}{0pt}
% \end{tabular}%
% }
% \flushleft{\footnotesize $^{*}p < 0.05$, $^{**}p < 0.01$, $^{***}p < 0.001$}
% \end{table}

\subsubsection{Size Bias}
\begin{figure}[htbp]
  \centering
  \includegraphics[width=\columnwidth]{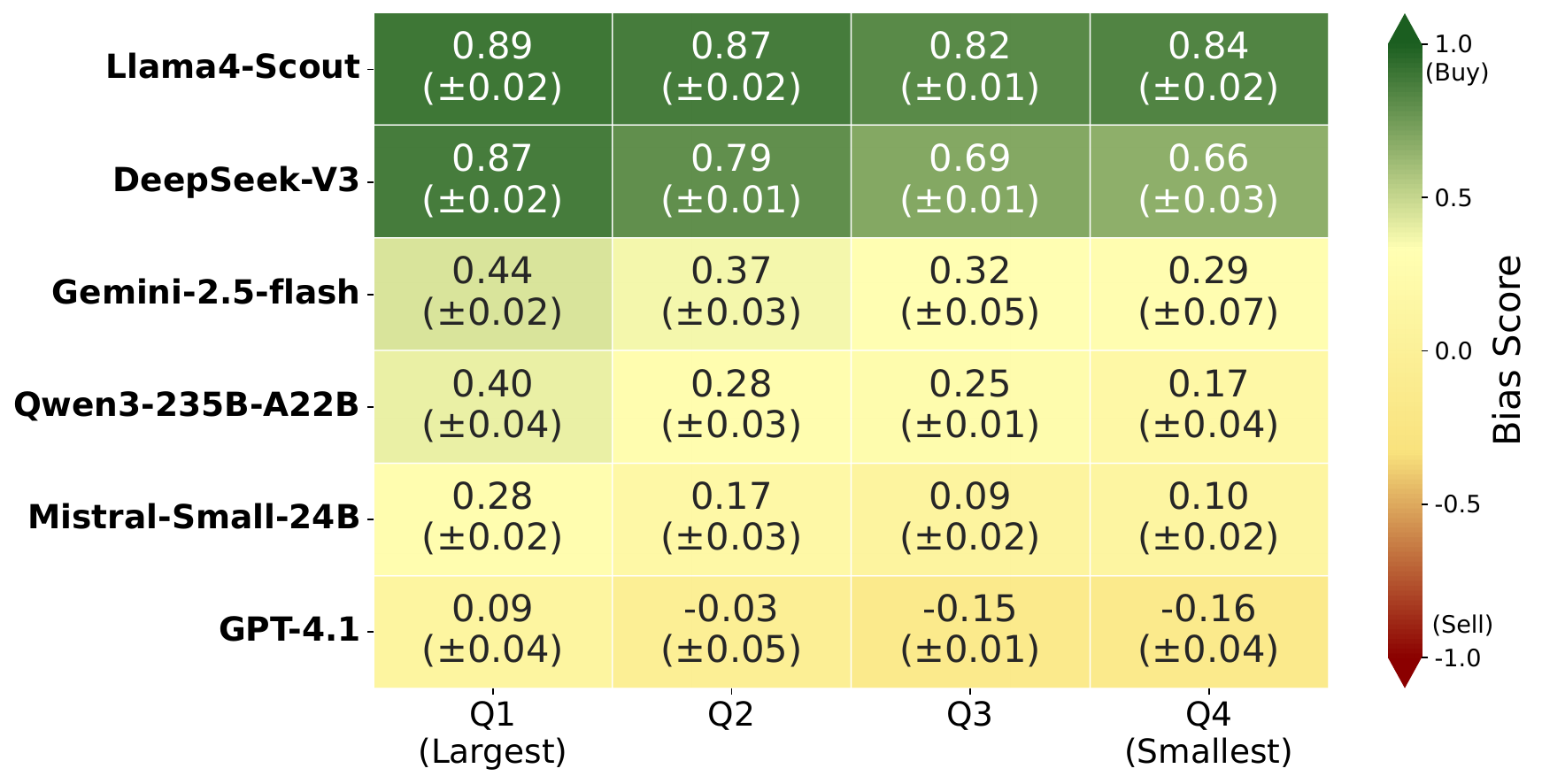}
  \caption{Size bias scores for each evaluated LLM across four market-capitalization quantiles (Q1: largest, Q4: smallest). Bias scores consistently decline as company size decreases.}
  \label{fig:size_heatmap}
\end{figure}

We investigate whether LLMs exhibit a bias for companies of a certain size, a factor that could influence their outputs in financial applications. To this end, we measure model bias scores across four market capitalization quantiles (Q1: highest, Q4: lowest), with detailed results presented in Figure \ref{fig:size_heatmap}. Across all models examined, we observe a consistent pattern: bias scores generally decline as company size decreases, indicating a stronger preference for large-capitalization firms that weakens toward smaller ones. Notably, one model’s scores even become negative in the lowest two quantiles (Q3 and Q4), suggesting not merely a loss of preference but at times even a negative stance toward smaller-capitalization companies. These results highlight that size bias is a pervasive feature among LLMs, and its intensity and direction can differ substantially depending on the model.

To statistically validate these observed trends, we performed independent-samples t-tests comparing the bias scores between the highest- and lowest-bias quantiles for each model (Table~\ref{tab:ttest_size}). The results show that the gap in bias scores is statistically significant across all models evaluated, confirming that size-related bias is a consistent characteristic of LLMs. Although the magnitude varies, every model shows a clear size-related difference in bias scores.

\begin{table}[htbp]
\caption{Independent-samples t-test results for the score gap between each model’s highest- and lowest-scoring quantiles.}
\label{tab:ttest_size}
\centering
\resizebox{0.9\columnwidth}{!}{% 
\begin{tabular}{lccc}
\specialrule{1.1pt}{0pt}{0pt}
\textbf{Model} & \textbf{High-Score} & \textbf{Low-Score} & \textbf{p-value} \\
\specialrule{1pt}{0pt}{0pt}
\texttt{Llama4-Scout} & Q1 & Q3 & $<0.001^{***}$ \\
\texttt{DeepSeek-V3} & Q1 (Largest) & Q4 (Smallest) & $<0.001^{***}$ \\
\texttt{Gemini-2.5-flash} & Q1 & Q4 & $<0.001^{***}$ \\
\texttt{Qwen3-235B} & Q1 & Q4 & $<0.001^{***}$ \\
\texttt{Mistral-Small} & Q1 & Q3 & $<0.001^{***}$ \\
\texttt{GPT-4.1} & Q1 & Q4 & $<0.001^{***}$ \\
\specialrule{1.1pt}{0pt}{0pt}
\end{tabular}%
}
\end{table}

% \begin{table}[htbp]
% \caption{Independent samples t-test of the preference gap between each model's highest and lowest preference quantiles.}
% \label{tab:ttest_size}
% \centering
% \resizebox{\columnwidth}{!}{%
% \begin{tabular}{lcccc}
% \specialrule{1.1pt}{0pt}{0pt}
% \textbf{Model} & \textbf{High-Pref} & \textbf{Low-Pref} & \textbf{Diff} & \textbf{p-value} \\
% \specialrule{1pt}{0pt}{0pt}
% \texttt{Llama4-Scout} & Q2 & Q3 & 0.0719 & 0.015$^{*}$ \\
% \texttt{DeepSeek-V3} & Q1 & Q4 & 0.1869 & $<0.001^{***}$ \\
% \texttt{Qwen3-235B} & Q1 & Q4 & 0.1178 & 0.004$^{**}$ \\
% \texttt{Gemini-2.5} & Q1 & Q3 & 0.1514 & $<0.001^{***}$ \\
% \texttt{GPT-4.1} & Q2 & Q4 & 0.0321 & 0.417 \\
% \texttt{Mistral-24B} & Q1 & Q4 & 0.0785 & 0.054 \\
% \specialrule{1.1pt}{0pt}{0pt}
% \end{tabular}%
% }
% \end{table}
We attribute these biases to underlying \textit{popularity effects}, wherein sectors and companies with greater market prominence—such as well-known industries or large-cap firms—are likely to appear more frequently and in richer contexts within the training corpora. As a result, LLMs tend to develop stronger priors for these dominant sectors and larger companies. This dual bias toward popular sectors and high-capitalization stocks has critical implications for financial applications: it can lead models to systematically overvalue certain industries and large-cap firms while undervaluing less represented sectors or smaller-cap companies, regardless of their fundamental merits. Practitioners should remain mindful of these tendencies and implement appropriate auditing or corrective measures when deploying LLMs for tasks such as portfolio construction.

% We attribute this behavior to a \textit{popularity effect}, wherein greater data volume and richness for larger, well-known corporations in the training corpora lead the models to develop stronger priors for them. This finding has critical implications for the application of LLMs in finance. The models' inherent inclination towards large-cap stocks could lead to the systematic overlooking of smaller-cap companies, irrespective of their fundamental merits. Therefore, we advise practitioners to be mindful of this characteristic and to account for it when using these models for tasks like portfolio construction.

\subsubsection{Momentum Bias}
In investment strategies, the momentum view involves favoring assets with recent strong performance, expecting trend continuation. In contrast, the contrarian view entails selecting underperforming assets in anticipation of mean reversion.

\begin{figure}[htbp]
    \centering
    \includegraphics[width=\columnwidth]{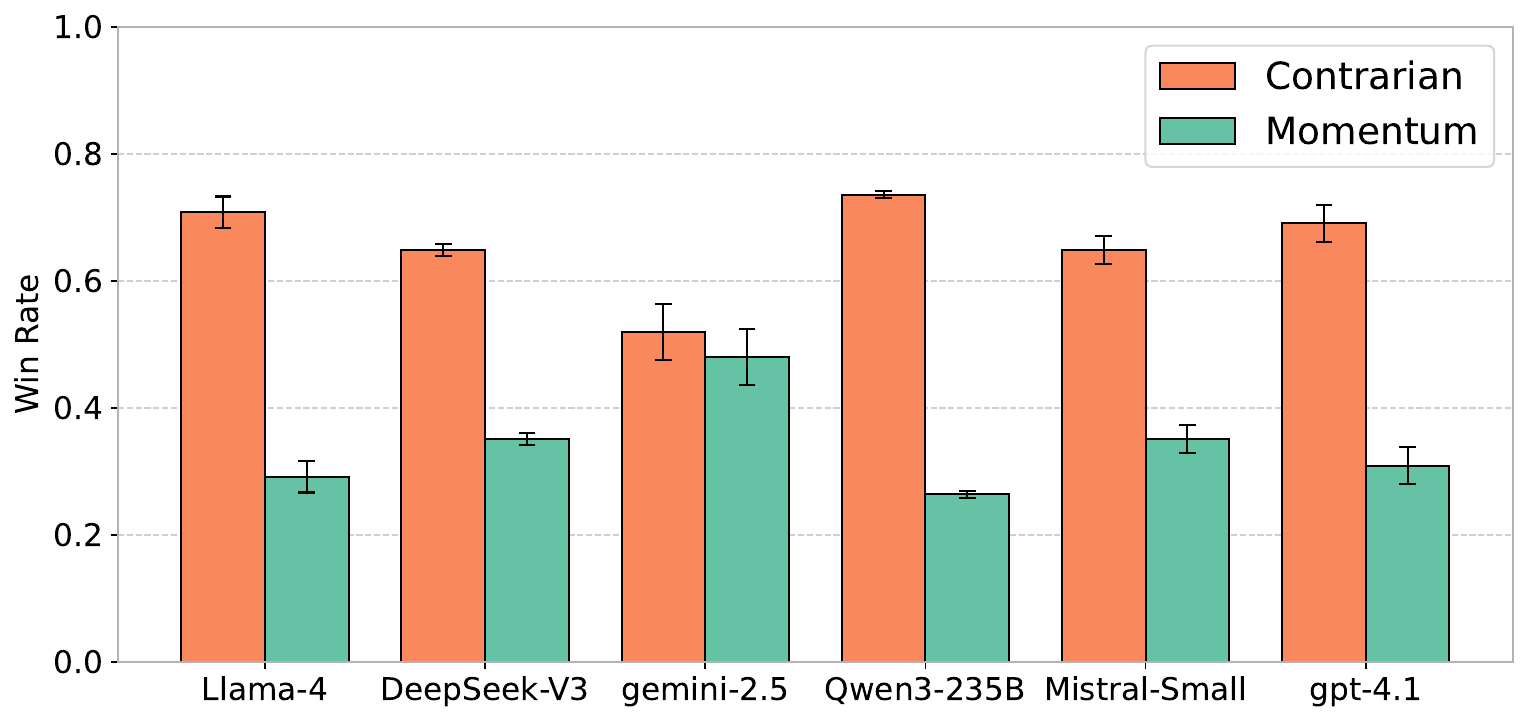}
    \caption{Win rates for Contrarian versus Momentum preferences for each model. The results show a consistent preference for the Contrarian view across most models.}
    \label{fig:view_preferences}
\end{figure}

Measuring model bias for investment styles like momentum or contrarian requires a different approach than the previously discussed sector or size analyses. Unlike a specific sector or size, an investment style can be explicitly framed as $\mathcal{E}_{\text{buy}}^{(s)}$ or $\mathcal{E}_{\text{sell}}^{(s)}$. We leverage this by designing a prompt where $\mathcal{E}_{\text{buy}}^{(s)}$ is based on one view (e.g., momentum), while $\mathcal{E}_{\text{sell}}^{(s)}$ is based on the opposing view (e.g., contrarian). To mitigate potential positional bias, we ensured a balanced experimental design where each view was used to generate both $\mathcal{E}_{\text{buy}}^{(s)}$ and $\mathcal{E}_{\text{sell}}^{(s)}$ an equal number of times.

In this setup, if the model ultimately chooses the buy action, the investment view that generated $\mathcal{E}_{\text{buy}}^{(s)}$ is considered to have won. We quantify the model's bias by repeating this process and calculating the win rate for each investment view.

\begin{figure*}[t!]
    \centering
    \includegraphics[width=\textwidth]{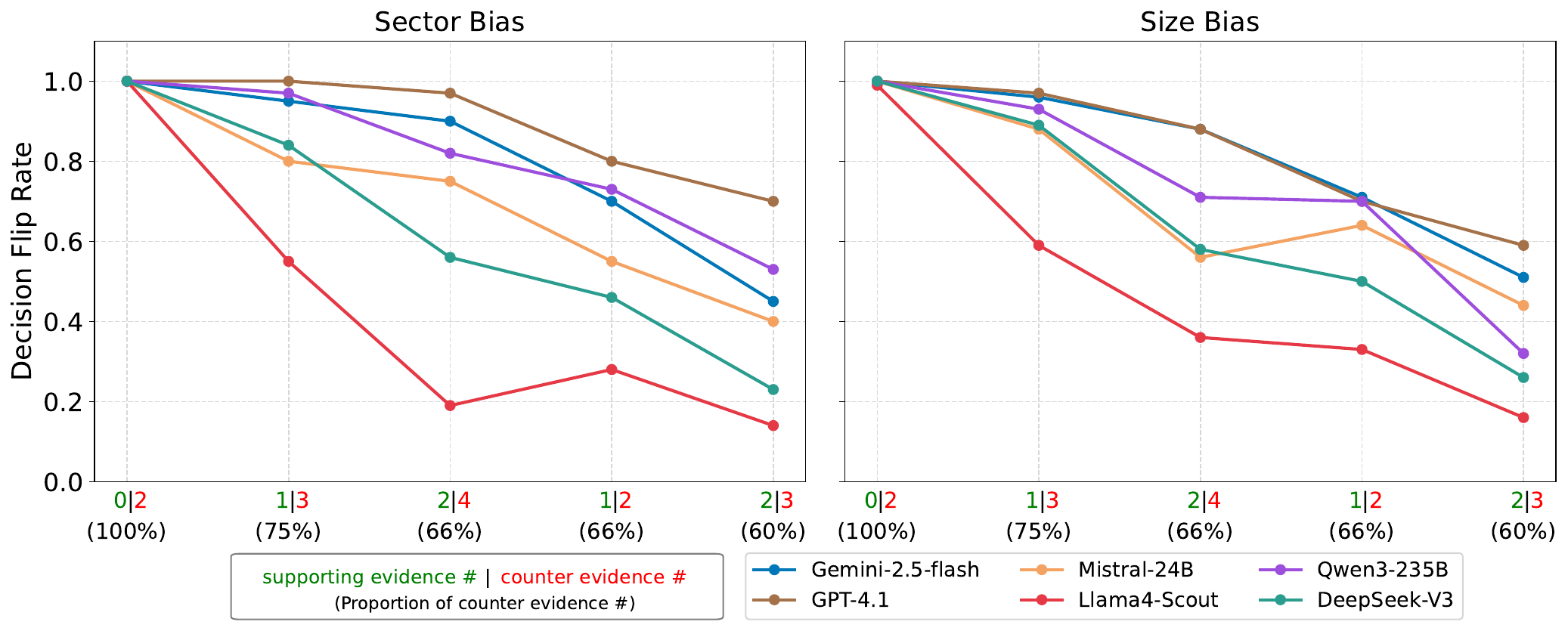}
    \caption{Decision flip rates under varying volumes of evidence for sector and size bias. The ratios (e.g., 2|3) denote the count of supporting vs. counter-
evidence, while the percentages represent the proportion of counter-evidence in the mix. The results reveal a sharp contrast:
the flip rate is near 1.0 for all models when only counter-evidence is presented (100\% case), but drops significantly the moment
any supporting evidence is introduced, indicating a strong difficulty in reversing decisions under conflicting information.}
    \label{fig:decision_flip}
\end{figure*}

Figure~\ref{fig:view_preferences} illustrates the bias of various models for contrarian versus momentum views. Our analysis reveals a consistent preference across all evaluated models toward the contrarian view. In particular, \texttt{Qwen3-235B} exhibits the strongest preference, with a high win rate for a contrarian stance and a correspondingly low rate for momentum. Models such as \texttt{DeepSeek-V3}, \texttt{Llama4-Scout}, and \texttt{GPT-4.1} also display clear contrarian inclinations, albeit with varying intensities. For \texttt{Gemini-2.5-flash}, the contrarian preference is evident but marginal, with win rates showing a negligible difference between the two views.

To statistically validate these observed tendencies, we performed a Chi-Square test to assess whether the difference in win rates between the contrarian and momentum views was significant (Table~\ref{tab:chi2_momentum}). The results show that all models exhibit a statistically significant preference for the contrarian view, confirming that this tendency is consistent across the evaluated LLMs.

\begin{table}[htbp]
\caption{Chi-Square test results for the score gap between contrarian and momentum views.}
\label{tab:chi2_momentum}
\centering
\resizebox{0.9\columnwidth}{!}{% 
\begin{tabular}{lccc}
\specialrule{1.1pt}{0pt}{0pt}
\textbf{Model} & \textbf{High-Score} & \textbf{Low-Score} & \textbf{p-value} \\
\specialrule{1pt}{0pt}{0pt}
\texttt{Llama4-Scout}   & contrarian & momentum & $<0.001^{***}$ \\
\texttt{DeepSeek-V3}    & contrarian & momentum & $<0.001^{***}$ \\
\texttt{Gemini-2.5-flash}     & contrarian & momentum & $0.0052^{**}$ \\
\texttt{Qwen3-235B}     & contrarian & momentum & $<0.001^{***}$ \\
\texttt{Mistral-Small}    & contrarian & momentum & $<0.001^{***}$ \\
\texttt{GPT-4.1}        & contrarian & momentum & $<0.001^{***}$ \\
\specialrule{1.1pt}{0pt}{0pt}
\end{tabular}%
}
\end{table}

These results highlight the importance of accounting for intrinsic view preferences in LLMs. A consistent bias toward the contrarian style can skew investment decisions, favoring underperforming assets even when market signals support momentum strategies. Recognizing these biases is essential for deploying LLMs in finance, where unmitigated view preferences may lead to unintended portfolio imbalances. Practitioners should audit models for these tendencies and consider mitigation strategies to ensure decisions align with intended investment objectives.

% These results highlight the need to account for intrinsic view preferences in LLMs, as they may cause discrepancies between anticipated and actual outputs in decision-making tasks.

\subsection{Bias Verification with Counter-Evidence}
\label{subsec:bias_verification}
This section examines the Decision Flip Rate, a metric quantifying the resilience of a model's initial bias when exposed to a high proportion of counter-evidence. The experiments aim to assess the resilience of this bias, with reported values indicating the frequency of decision reversals under controlled conditions where evidence is intentionally skewed against the model's established bias.

\subsubsection{Approach 1: Verification by Evidence Volume}
This experiment was designed to observe how much an initial decision is reversed when a model is provided with a weighted amount of evidence that opposes its existing bias. This rate of change, measured as $\phi_{s^*}^{\text{vol}}$, probes the persistence of bias, yielding results consistent with prior observations of contradictory LLM behaviors~\cite{xie2023adaptive}. Figure~\ref{fig:decision_flip} presents the $\phi_{s^*}^{\text{vol}}$ values for each model at various ratios of supporting versus counter-evidence (e.g., 2|3).

When provided only with counter-evidence, all models exhibited high receptivity, overriding their internal knowledge and achieving $\phi_{s^*}^{\text{vol}}$ values near 1.0. However, in situations where supporting and counter-evidence were mixed, creating a knowledge conflict, the $\phi_{s^*}^{\text{vol}}$ values dropped sharply. This phenomenon occurred despite the amount of counter-evidence always being greater than the supporting evidence in all experimental conditions, strongly suggesting that models selectively adhere to information that aligns with their pre-existing inclinations.

This rigidity was more evident in models with strong inherent bias. For instance, \texttt{Llama4-Scout} and \texttt{DeepSeek-V3}, which had high bias scores across sectors, recorded particularly low $\phi_{s^*}^{\text{vol}}$ values. These models struggled to reverse their decisions, especially when the volume difference between supporting and counter-evidence was small. Similarly, \texttt{Qwen3-235b} also showed reduced flexibility at lower proportions of counter-evidence.

In contrast, models with overall lower bias scores demonstrated greater adaptability. \texttt{GPT-4.1} and \texttt{Gemini-2.5-flash} maintained higher $\phi_s^{\text{vol}}$ values, remaining relatively responsive even when the difference in evidence volume was minimal. Although their $\phi_{s^*}^{\text{vol}}$ values fell short of expectations despite the counter-evidence majority, this pattern shows a direct correlation with initial bias strength. In other words, the stronger a model's inherent bias, the more its stubbornness is amplified when the difference in the volume of supporting and counter-evidence is small. Consequently, this finding suggests a significant risk in real-world financial contexts where conflicting information is present (for instance, when price indicators are negative but related news is positive). In such cases, a model could trust only one side of the evidence due to its inherent bias, leading to flawed judgments.

\subsubsection{Approach 2: Verification by Evidence Intensity}
\begin{figure}[htbp]
    \centering
    \includegraphics[width=\columnwidth]{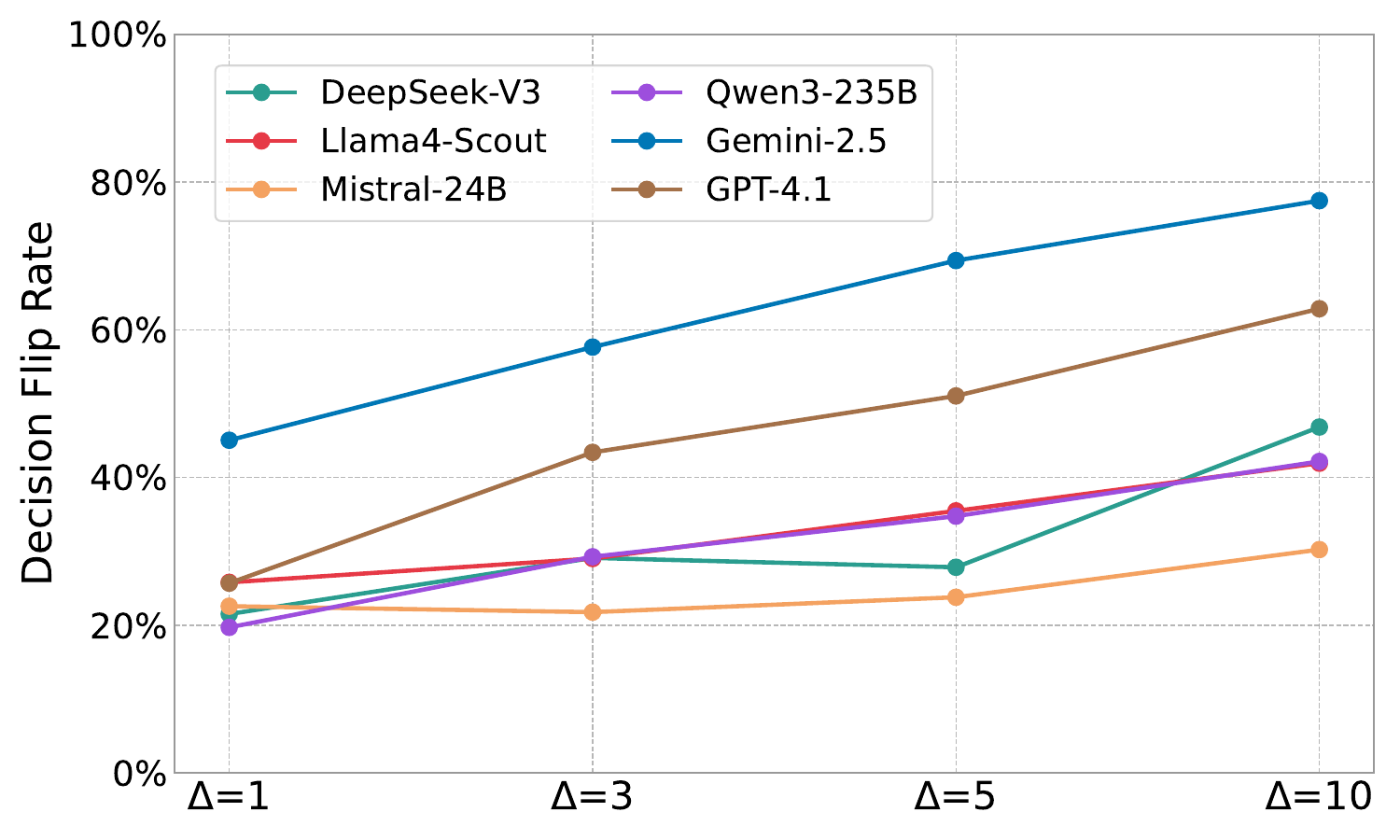}
    \caption{Decision flip rates under varying intensities of evidence for momentum bias. Even as counter-evidence intensity ($\Delta$) increases, the decision flip rate ($\phi_s^{\text{int}}$) for most models remains low, indicating strong confirmation bias. \texttt{Gemini-2.5-flash}, which had the least initial bias, shows the most flexible response.}
    \label{fig:mc_change}
\end{figure}

This approach investigates model sensitivity by maintaining volumetric parity while escalating the intensity increment, $\Delta$, of the counter-evidence. Figure~\ref{fig:mc_change} plots the intensity-driven flip rate ($\phi_{s^*}^{\text{int}}$) against $\Delta$ values of 1, 3, 5, and 10. The results delineate a clear sensitivity spectrum among the models, which correlates with the prior view bias analysis.

While the graph shows a gradual upward trend in $\phi_{s^*}^{\text{int}}$ for all models as $\Delta$ increases, the more notable finding lies in the magnitude of this increase and the final values. Even when presented with very strong counter-evidence ($\Delta=10$), the majority of models recorded low $\phi_{s^*}^{\text{int}}$ values below 60\%. This signifies that the models' confirmation bias is not easily overcome, even by qualitatively superior counter-evidence.

Finally, we observe a strong correlation between bias strength and model rigidity. Notably, \texttt{Gemini-2.5-flash}, the model with the most balanced bias profile, achieves a substantially higher flip rate ($\phi_{s^*}^{\text{int}}$) than its counterparts. This finding suggests that a low initial bias is a key predictor of a model's ability to adapt when presented with qualitatively superior counter-evidence.

Conversely, models identified with stronger and more polarized bias formed the lower-performing group. These models consistently recorded low $\phi_{s^*}^{\text{int}}$ values, signifying a more stubborn confirmation bias. The behavior of \texttt{Qwen3-235B}, which had one of the largest bias gaps, exemplifies this resistance, as it remains one of the least likely models to reverse its decision even when the counter-evidence is significantly more intense.

Synthesizing these results provides a deeper insight into model behavior. Even when presented with qualitatively superior counter-evidence ($\Delta=10$), models show a strong tendency to struggle with decision reversal due to their bias. This rigidity poses a tangible risk when considering the findings from our prior analysis, where all models commonly biased a contrarian view over a momentum view. It implies that a model's inherent bias toward a specific investment perspective could cause it to ignore or undervalue strong opposing evidence, potentially leading to skewed conclusions.

\subsection{Decision Uncertainty}
\label{subsec:uncertainty}
\begin{figure}[htbp]
    \centering
    \includegraphics[width=\columnwidth]{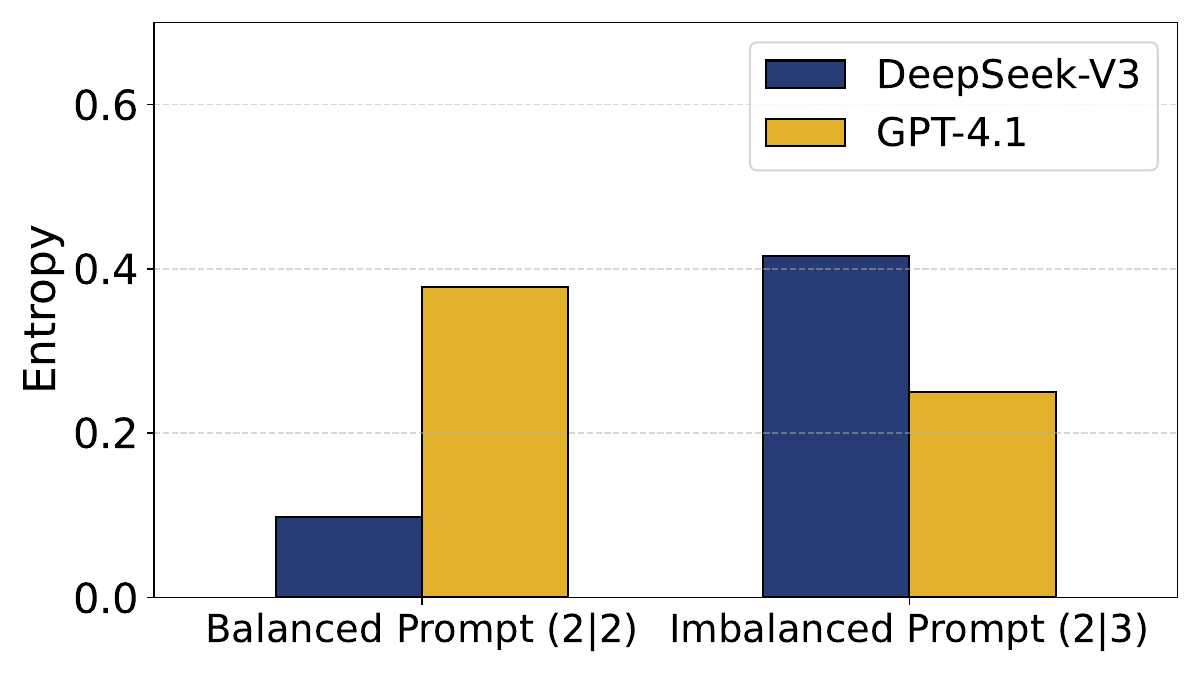}
    \caption{Entropy comparison between a high-score model (\texttt{DeepSeek-V3}) and a low-score model (\texttt{GPT-4.1}). The pattern inverts when the prompt shifts from balanced to imbalanced.}
    \label{fig:figure6}
\end{figure}

To quantify the internal uncertainty experienced by the model, we conducted an entropy analysis (Figure \ref{fig:figure6}). The uncertainty was assessed using the Shannon entropy, $H$, computed directly from the probability distribution the model assigned over the potential action tokens during generation. Specifically, letting $P(\text{buy})$ and $P(\text{sell})$ represent the probabilities assigned by the model to the respective action tokens, entropy is formally defined as:
\begin{equation}
\label{eq:shannon_entropy}
H(\text{Decision}) = -\sum_{x \in \{\text{buy, sell}\}} P(x) \log_2 P(x).
\end{equation}
A higher entropy value indicates greater uncertainty, while a lower entropy corresponds to higher confidence in the decision-making process. This analysis compares the uncertainty between two distinct models: \texttt{DeepSeek-V3}, selected as a representative for its high bias score, and \texttt{GPT-4.1}, chosen for its low bias score.

Under the Balanced Prompt condition, where evidence was presented in equilibrium, the two models exhibited distinct entropy patterns. \texttt{GPT-4.1}, characterized by its low bias score, recorded high entropy, signifying a state of high uncertainty and an inability to commit to a decision. Conversely, \texttt{DeepSeek-V3}, with its strong bias, showed very low entropy. This suggests its internal references easily broke the tie presented by the external evidence, allowing it to make a confident decision.

Interestingly, under the Imbalanced Prompt condition, where more counter-evidence was presented, the entropy pattern inverted. The entropy of \texttt{DeepSeek-V3} rose sharply, suggesting it was experiencing cognitive dissonance from the conflict between its strong bias and the clear external counter-evidence. In contrast, the entropy of \texttt{GPT-4.1} decreased. With less internal bias, it could confidently align with the majority of the evidence, which resolved its uncertainty from the previous condition.

Ultimately, stronger bias appears to intensify hesitation and uncertainty when confronted with conflicting external evidence. The entropy analysis thus highlights how bias affects not only the direction of decisions but also the confidence levels and internal cognitive conflicts experienced by models during decision-making.

\section{Limitations}
This study has several limitations. First, our evidence was generated by a single LLM and simplified to a numerical value, an approach that may not fully capture the biases of the generator model or the complexity of real-world information. Second, the experimental design is limited in verifying reasoning model bias, as models can bypass the intended conflict by recognizing the evidence imbalance. Third, our analysis is static, providing a snapshot of model biases at one point in time without capturing their temporal dynamics.

\section{Conclusion}
This study systematically investigated the latent biases of LLMs in financial contexts and analyzed the critical issues, such as confirmation bias, that arise under informational conflict. We sought to answer two key research questions: what financial factor biases LLMs exhibit, and what problems these biases cause when the models are forced to make decisions with conflicting evidence. The results demonstrate that LLMs are not neutral decision-makers, possessing distinct biases for factors like sector, size, and momentum. While the strength of bias was highly model-dependent, a common inclination towards the Technology sector was observed, alongside a shared bias towards large-size stocks and a consistent contrarian bias over a momentum-based investment style.

Our bias verification experiments revealed that these latent biases directly lead to significant confirmation bias when challenged. While the models correctly reversed their decisions when faced only with counter-evidence, their objectivity sharply decreased in scenarios with mixed, conflicting arguments. This stubbornness was particularly pronounced in models that initially exhibited stronger biases, demonstrating a clear link between the intensity of a latent bias and the severity of the resulting confirmation bias. Furthermore, an entropy analysis quantified the models' internal uncertainty, showing that models with strong initial biases experience greater cognitive conflict when their established views are challenged by contradictory facts.

These findings have significant implications for the financial industry, where sound decisions require a holistic evaluation of all available information. The inherent biases of LLMs can distort this crucial process, compromising the reliability of AI-driven financial services. When a model's opaque biases conflict with user intent, its judgments become distorted and unpredictable. By illuminating how latent biases escalate into such unreliable outcomes, this study represents a critical step toward building more transparent, predictable, and ultimately, \textbf{Trustworthy AI} for finance. 
% Future work should focus on developing mitigation techniques to neutralize these identified biases.

\begin{acks}
This work was supported by National Research Foundation of Korea (NRF) grant (No. NRF-2022R1I1A4069163) and Institute of Information \& Communications Technology Planning \& Evaluation(IITP) grant (No. RS-2020-II201336, Artificial Intelligence Graduate School Program(UNIST)) funded by the Korea government(MSIT).
\end{acks}

\bibliographystyle{ACM-Reference-Format}
\bibliography{ref}


\begin{thebibliography}{31}

%%% the Web address.

\ifx \showCODEN    \undefined \def \showCODEN     #1{\unskip}     \fi
\ifx \showDOI      \undefined \def \showDOI       #1{#1}\fi
\ifx \showISBNx    \undefined \def \showISBNx     #1{\unskip}     \fi
\ifx \showISBNxiii \undefined \def \showISBNxiii  #1{\unskip}     \fi
\ifx \showISSN     \undefined \def \showISSN      #1{\unskip}     \fi
\ifx \showLCCN     \undefined \def \showLCCN      #1{\unskip}     \fi
\ifx \shownote     \undefined \def \shownote      #1{#1}          \fi
\ifx \showarticletitle \undefined \def \showarticletitle #1{#1}   \fi
\ifx \showURL      \undefined \def \showURL       {\relax}        \fi
% invisible to TeX
\providecommand\bibfield[2]{#2}
\providecommand\bibinfo[2]{#2}
\providecommand\natexlab[1]{#1}
\providecommand\showeprint[2][]{arXiv:#2}

\bibitem[Achiam et~al\mbox{.}(2023)]%
        {achiam2023gpt}
\bibfield{author}{\bibinfo{person}{Josh Achiam}, \bibinfo{person}{Steven Adler}, \bibinfo{person}{Sandhini Agarwal}, \bibinfo{person}{Lama Ahmad}, \bibinfo{person}{Ilge Akkaya}, \bibinfo{person}{Florencia~Leoni Aleman}, \bibinfo{person}{Diogo Almeida}, \bibinfo{person}{Janko Altenschmidt}, \bibinfo{person}{Sam Altman}, \bibinfo{person}{Shyamal Anadkat}, {et~al\mbox{.}}} \bibinfo{year}{2023}\natexlab{}.
\newblock \showarticletitle{Gpt-4 technical report}.
\newblock \bibinfo{journal}{\emph{arXiv preprint arXiv:2303.08774}} (\bibinfo{year}{2023}).
\newblock


\bibitem[Bini et~al\mbox{.}(2025)]%
        {bini2025behavioral}
\bibfield{author}{\bibinfo{person}{Pietro Bini}, \bibinfo{person}{Lin~William Cong}, \bibinfo{person}{Xing Huang}, {and} \bibinfo{person}{Lawrence~J Jin}.} \bibinfo{year}{2025}\natexlab{}.
\newblock \showarticletitle{Behavioral Economics of AI: LLM Biases and Corrections}.
\newblock \bibinfo{journal}{\emph{Available at SSRN 5213130}} (\bibinfo{year}{2025}).
\newblock


\bibitem[Cao et~al\mbox{.}(2025)]%
        {cao2025llms}
\bibfield{author}{\bibinfo{person}{Sean Cao}, \bibinfo{person}{Charles~CY Wang}, {and} \bibinfo{person}{Yi Xiang}.} \bibinfo{year}{2025}\natexlab{}.
\newblock \showarticletitle{When LLMs Go Abroad: Foreign Bias in AI Financial Predictions}.
\newblock \bibinfo{journal}{\emph{Available at SSRN 5440116}} (\bibinfo{year}{2025}).
\newblock


\bibitem[Chen et~al\mbox{.}(2024)]%
        {chen2024what}
\bibfield{author}{\bibinfo{person}{Shuaiyu Chen}, \bibinfo{person}{T~Clifton Green}, \bibinfo{person}{Huseyin Gulen}, {and} \bibinfo{person}{Dexin Zhou}.} \bibinfo{year}{2024}\natexlab{}.
\newblock \showarticletitle{What does chatgpt make of historical stock returns? extrapolation and miscalibration in llm stock return forecasts}.
\newblock \bibinfo{journal}{\emph{arXiv preprint arXiv:2409.11540}} (\bibinfo{year}{2024}).
\newblock


\bibitem[Comanici et~al\mbox{.}(2025)]%
        {comanici2025gemini}
\bibfield{author}{\bibinfo{person}{Gheorghe Comanici}, \bibinfo{person}{Eric Bieber}, \bibinfo{person}{Mike Schaekermann}, \bibinfo{person}{Ice Pasupat}, \bibinfo{person}{Noveen Sachdeva}, \bibinfo{person}{Inderjit Dhillon}, \bibinfo{person}{Marcel Blistein}, \bibinfo{person}{Ori Ram}, \bibinfo{person}{Dan Zhang}, \bibinfo{person}{Evan Rosen}, {et~al\mbox{.}}} \bibinfo{year}{2025}\natexlab{}.
\newblock \showarticletitle{Gemini 2.5: Pushing the Frontier with Advanced Reasoning, Multimodality, Long Context, and Next Generation Agentic Capabilities}.
\newblock \bibinfo{journal}{\emph{arXiv preprint arXiv:2507.06261}} (\bibinfo{year}{2025}).
\newblock


\bibitem[Cook and Kazinnik(2025)]%
        {cook2025social}
\bibfield{author}{\bibinfo{person}{Thomas~R Cook} {and} \bibinfo{person}{Sophia Kazinnik}.} \bibinfo{year}{2025}\natexlab{}.
\newblock \showarticletitle{Social Group Bias in AI Finance}.
\newblock \bibinfo{journal}{\emph{arXiv preprint arXiv:2506.17490}} (\bibinfo{year}{2025}).
\newblock


\bibitem[Echterhoff et~al\mbox{.}(2024)]%
        {echterhoff2024cognitive}
\bibfield{author}{\bibinfo{person}{Jessica Echterhoff}, \bibinfo{person}{Yao Liu}, \bibinfo{person}{Abeer Alessa}, \bibinfo{person}{Julian McAuley}, {and} \bibinfo{person}{Zexue He}.} \bibinfo{year}{2024}\natexlab{}.
\newblock \showarticletitle{Cognitive bias in decision-making with LLMs}.
\newblock \bibinfo{journal}{\emph{arXiv preprint arXiv:2403.00811}} (\bibinfo{year}{2024}).
\newblock


\bibitem[Ferrando et~al\mbox{.}(2024)]%
        {ferrando2024know}
\bibfield{author}{\bibinfo{person}{Javier Ferrando}, \bibinfo{person}{Oscar Obeso}, \bibinfo{person}{Senthooran Rajamanoharan}, {and} \bibinfo{person}{Neel Nanda}.} \bibinfo{year}{2024}\natexlab{}.
\newblock \showarticletitle{Do i know this entity? knowledge awareness and hallucinations in language models}.
\newblock \bibinfo{journal}{\emph{arXiv preprint arXiv:2411.14257}} (\bibinfo{year}{2024}).
\newblock


\bibitem[Hwang et~al\mbox{.}(2025)]%
        {hwang2025decision}
\bibfield{author}{\bibinfo{person}{Yoontae Hwang}, \bibinfo{person}{Yaxuan Kong}, \bibinfo{person}{Stefan Zohren}, {and} \bibinfo{person}{Yongjae Lee}.} \bibinfo{year}{2025}\natexlab{}.
\newblock \showarticletitle{Decision-informed neural networks with large language model integration for portfolio optimization}.
\newblock \bibinfo{journal}{\emph{arXiv preprint arXiv:2502.00828}} (\bibinfo{year}{2025}).
\newblock


\bibitem[Jin et~al\mbox{.}(2024)]%
        {jin2024tug}
\bibfield{author}{\bibinfo{person}{Zhuoran Jin}, \bibinfo{person}{Pengfei Cao}, \bibinfo{person}{Yubo Chen}, \bibinfo{person}{Kang Liu}, \bibinfo{person}{Xiaojian Jiang}, \bibinfo{person}{Jiexin Xu}, \bibinfo{person}{Qiuxia Li}, {and} \bibinfo{person}{Jun Zhao}.} \bibinfo{year}{2024}\natexlab{}.
\newblock \showarticletitle{Tug-of-war between knowledge: Exploring and resolving knowledge conflicts in retrieval-augmented language models}.
\newblock \bibinfo{journal}{\emph{arXiv preprint arXiv:2402.14409}} (\bibinfo{year}{2024}).
\newblock


\bibitem[Kim et~al\mbox{.}(2023)]%
        {kim2023llms}
\bibfield{author}{\bibinfo{person}{Seonmi Kim}, \bibinfo{person}{Seyoung Kim}, \bibinfo{person}{Yejin Kim}, \bibinfo{person}{Junpyo Park}, \bibinfo{person}{Seongjin Kim}, \bibinfo{person}{Moolkyeol Kim}, \bibinfo{person}{Chang~Hwan Sung}, \bibinfo{person}{Joohwan Hong}, {and} \bibinfo{person}{Yongjae Lee}.} \bibinfo{year}{2023}\natexlab{}.
\newblock \showarticletitle{LLMs analyzing the analysts: Do BERT and GPT extract more value from financial analyst reports?}. In \bibinfo{booktitle}{\emph{Proceedings of the Fourth ACM International Conference on AI in Finance}}. \bibinfo{pages}{383--391}.
\newblock


\bibitem[Ko and Lee(2024)]%
        {ko2024can}
\bibfield{author}{\bibinfo{person}{Hyungjin Ko} {and} \bibinfo{person}{Jaewook Lee}.} \bibinfo{year}{2024}\natexlab{}.
\newblock \showarticletitle{Can ChatGPT improve investment decisions? From a portfolio management perspective}.
\newblock \bibinfo{journal}{\emph{Finance Research Letters}}  \bibinfo{volume}{64} (\bibinfo{year}{2024}), \bibinfo{pages}{105433}.
\newblock


\bibitem[Kumaran et~al\mbox{.}(2025)]%
        {kumaran2025overconfidence}
\bibfield{author}{\bibinfo{person}{Dharshan Kumaran}, \bibinfo{person}{Stephen~M Fleming}, \bibinfo{person}{Larisa Markeeva}, \bibinfo{person}{Joe Heyward}, \bibinfo{person}{Andrea Banino}, \bibinfo{person}{Mrinal Mathur}, \bibinfo{person}{Razvan Pascanu}, \bibinfo{person}{Simon Osindero}, \bibinfo{person}{Benedetto De~Martino}, \bibinfo{person}{Petar Velickovic}, {et~al\mbox{.}}} \bibinfo{year}{2025}\natexlab{}.
\newblock \showarticletitle{How Overconfidence in Initial Choices and Underconfidence Under Criticism Modulate Change of Mind in Large Language Models}.
\newblock \bibinfo{journal}{\emph{arXiv preprint arXiv:2507.03120}} (\bibinfo{year}{2025}).
\newblock


\bibitem[Lee et~al\mbox{.}(2025)]%
        {lee2025integrating}
\bibfield{author}{\bibinfo{person}{Youngbin Lee}, \bibinfo{person}{Yejin Kim}, \bibinfo{person}{Suin Kim}, {and} \bibinfo{person}{Yongjae Lee}.} \bibinfo{year}{2025}\natexlab{}.
\newblock \showarticletitle{Integrating LLM-Generated Views into Mean-Variance Optimization Using the Black-Litterman Model}.
\newblock \bibinfo{journal}{\emph{arXiv preprint arXiv:2504.14345}} (\bibinfo{year}{2025}).
\newblock


\bibitem[Liu et~al\mbox{.}(2024)]%
        {liu2024deepseek}
\bibfield{author}{\bibinfo{person}{Aixin Liu}, \bibinfo{person}{Bei Feng}, \bibinfo{person}{Bing Xue}, \bibinfo{person}{Bingxuan Wang}, \bibinfo{person}{Bochao Wu}, \bibinfo{person}{Chengda Lu}, \bibinfo{person}{Chenggang Zhao}, \bibinfo{person}{Chengqi Deng}, \bibinfo{person}{Chenyu Zhang}, \bibinfo{person}{Chong Ruan}, {et~al\mbox{.}}} \bibinfo{year}{2024}\natexlab{}.
\newblock \showarticletitle{Deepseek-v3 technical report}.
\newblock \bibinfo{journal}{\emph{arXiv preprint arXiv:2412.19437}} (\bibinfo{year}{2024}).
\newblock


\bibitem[Lo and Ross(2024)]%
        {lo2024can}
\bibfield{author}{\bibinfo{person}{Andrew~W Lo} {and} \bibinfo{person}{Jillian Ross}.} \bibinfo{year}{2024}\natexlab{}.
\newblock \showarticletitle{Can ChatGPT plan your retirement?: Generative AI and financial advice}.
\newblock \bibinfo{journal}{\emph{Generative AI and Financial Advice (February 11, 2024)}} (\bibinfo{year}{2024}).
\newblock


\bibitem[{Meta AI}(2025)]%
        {meta-llama4}
\bibfield{author}{\bibinfo{person}{{Meta AI}}.} \bibinfo{year}{2025}\natexlab{}.
\newblock \bibinfo{booktitle}{\emph{Llama 4}}.
\newblock
\urldef\tempurl%
\url{https://www.llama.com/models/llama-4/}
\showURL{%
\tempurl}
\newblock
\shownote{Accessed: 2025-05-07}.


\bibitem[{Mistral AI Team}(2025)]%
        {mistral-small3}
\bibfield{author}{\bibinfo{person}{{Mistral AI Team}}.} \bibinfo{year}{2025}\natexlab{}.
\newblock \bibinfo{booktitle}{\emph{Mistral Small 3}}.
\newblock
\urldef\tempurl%
\url{https://mistral.ai/news/mistral-small-3}
\showURL{%
\tempurl}
\newblock
\shownote{Accessed: 2025-01-30}.


\bibitem[Nakagawa et~al\mbox{.}(2024)]%
        {nakagawa2024evaluating}
\bibfield{author}{\bibinfo{person}{Kei Nakagawa}, \bibinfo{person}{Masanori Hirano}, {and} \bibinfo{person}{Yugo Fujimoto}.} \bibinfo{year}{2024}\natexlab{}.
\newblock \showarticletitle{Evaluating company-specific biases in financial sentiment analysis using large language models}. In \bibinfo{booktitle}{\emph{2024 IEEE International Conference on Big Data (BigData)}}. IEEE, \bibinfo{pages}{6614--6623}.
\newblock


\bibitem[Ouyang et~al\mbox{.}(2024)]%
        {ouyang2024risk}
\bibfield{author}{\bibinfo{person}{Shumiao Ouyang}, \bibinfo{person}{Hayong Yun}, {and} \bibinfo{person}{Xingjian Zheng}.} \bibinfo{year}{2024}\natexlab{}.
\newblock \showarticletitle{How ethical should AI be? how AI alignment shapes the risk preferences of llms}.
\newblock \bibinfo{journal}{\emph{arXiv preprint arXiv:2406.01168}} (\bibinfo{year}{2024}).
\newblock


\bibitem[Ross et~al\mbox{.}(2024)]%
        {ross2024llm}
\bibfield{author}{\bibinfo{person}{Jillian Ross}, \bibinfo{person}{Yoon Kim}, {and} \bibinfo{person}{Andrew~W Lo}.} \bibinfo{year}{2024}\natexlab{}.
\newblock \showarticletitle{LLM economicus? mapping the behavioral biases of LLMs via utility theory}.
\newblock \bibinfo{journal}{\emph{arXiv preprint arXiv:2408.02784}} (\bibinfo{year}{2024}).
\newblock


\bibitem[Stureborg et~al\mbox{.}(2024)]%
        {stureborg2024large}
\bibfield{author}{\bibinfo{person}{Rickard Stureborg}, \bibinfo{person}{Dimitris Alikaniotis}, {and} \bibinfo{person}{Yoshi Suhara}.} \bibinfo{year}{2024}\natexlab{}.
\newblock \showarticletitle{Large language models are inconsistent and biased evaluators}.
\newblock \bibinfo{journal}{\emph{arXiv preprint arXiv:2405.01724}} (\bibinfo{year}{2024}).
\newblock


\bibitem[Sun et~al\mbox{.}(2025)]%
        {sun2025seen}
\bibfield{author}{\bibinfo{person}{Kaiser Sun}, \bibinfo{person}{Fan Bai}, {and} \bibinfo{person}{Mark Dredze}.} \bibinfo{year}{2025}\natexlab{}.
\newblock \showarticletitle{What Is Seen Cannot Be Unseen: The Disruptive Effect of Knowledge Conflict on Large Language Models}.
\newblock \bibinfo{journal}{\emph{arXiv preprint arXiv:2506.06485}} (\bibinfo{year}{2025}).
\newblock


\bibitem[Tan et~al\mbox{.}(2024)]%
        {tan2024blinded}
\bibfield{author}{\bibinfo{person}{Hexiang Tan}, \bibinfo{person}{Fei Sun}, \bibinfo{person}{Wanli Yang}, \bibinfo{person}{Yuanzhuo Wang}, \bibinfo{person}{Qi Cao}, {and} \bibinfo{person}{Xueqi Cheng}.} \bibinfo{year}{2024}\natexlab{}.
\newblock \showarticletitle{Blinded by generated contexts: How language models merge generated and retrieved contexts when knowledge conflicts?}
\newblock \bibinfo{journal}{\emph{arXiv preprint arXiv:2401.11911}} (\bibinfo{year}{2024}).
\newblock


\bibitem[Xie et~al\mbox{.}(2023)]%
        {xie2023adaptive}
\bibfield{author}{\bibinfo{person}{Jian Xie}, \bibinfo{person}{Kai Zhang}, \bibinfo{person}{Jiangjie Chen}, \bibinfo{person}{Renze Lou}, {and} \bibinfo{person}{Yu Su}.} \bibinfo{year}{2023}\natexlab{}.
\newblock \showarticletitle{Adaptive chameleon or stubborn sloth: Revealing the behavior of large language models in knowledge conflicts}. In \bibinfo{booktitle}{\emph{The Twelfth International Conference on Learning Representations}}.
\newblock


\bibitem[Yang et~al\mbox{.}(2025)]%
        {yang2025qwen3}
\bibfield{author}{\bibinfo{person}{An Yang}, \bibinfo{person}{Anfeng Li}, \bibinfo{person}{Baosong Yang}, \bibinfo{person}{Beichen Zhang}, \bibinfo{person}{Binyuan Hui}, \bibinfo{person}{Bo Zheng}, \bibinfo{person}{Bowen Yu}, \bibinfo{person}{Chang Gao}, \bibinfo{person}{Chengen Huang}, \bibinfo{person}{Chenxu Lv}, {et~al\mbox{.}}} \bibinfo{year}{2025}\natexlab{}.
\newblock \showarticletitle{Qwen3 technical report}.
\newblock \bibinfo{journal}{\emph{arXiv preprint arXiv:2505.09388}} (\bibinfo{year}{2025}).
\newblock


\bibitem[Yu et~al\mbox{.}(2024)]%
        {yu2024fincon}
\bibfield{author}{\bibinfo{person}{Yangyang Yu}, \bibinfo{person}{Zhiyuan Yao}, \bibinfo{person}{Haohang Li}, \bibinfo{person}{Zhiyang Deng}, \bibinfo{person}{Yuechen Jiang}, \bibinfo{person}{Yupeng Cao}, \bibinfo{person}{Zhi Chen}, \bibinfo{person}{Jordan Suchow}, \bibinfo{person}{Zhenyu Cui}, \bibinfo{person}{Rong Liu}, {et~al\mbox{.}}} \bibinfo{year}{2024}\natexlab{}.
\newblock \showarticletitle{Fincon: A synthesized llm multi-agent system with conceptual verbal reinforcement for enhanced financial decision making}.
\newblock \bibinfo{journal}{\emph{Advances in Neural Information Processing Systems}}  \bibinfo{volume}{37} (\bibinfo{year}{2024}), \bibinfo{pages}{137010--137045}.
\newblock


\bibitem[Zhang et~al\mbox{.}(2024)]%
        {zhang2024multimodal}
\bibfield{author}{\bibinfo{person}{Wentao Zhang}, \bibinfo{person}{Lingxuan Zhao}, \bibinfo{person}{Haochong Xia}, \bibinfo{person}{Shuo Sun}, \bibinfo{person}{Jiaze Sun}, \bibinfo{person}{Molei Qin}, \bibinfo{person}{Xinyi Li}, \bibinfo{person}{Yuqing Zhao}, \bibinfo{person}{Yilei Zhao}, \bibinfo{person}{Xinyu Cai}, {et~al\mbox{.}}} \bibinfo{year}{2024}\natexlab{}.
\newblock \showarticletitle{A multimodal foundation agent for financial trading: Tool-augmented, diversified, and generalist}. In \bibinfo{booktitle}{\emph{Proceedings of the 30th acm sigkdd conference on knowledge discovery and data mining}}. \bibinfo{pages}{4314--4325}.
\newblock


\bibitem[Zhi et~al\mbox{.}(2025)]%
        {zhi2025exposing}
\bibfield{author}{\bibinfo{person}{Yuhan Zhi}, \bibinfo{person}{Xiaoyu Zhang}, \bibinfo{person}{Longtian Wang}, \bibinfo{person}{Shumin Jiang}, \bibinfo{person}{Shiqing Ma}, \bibinfo{person}{Xiaohong Guan}, {and} \bibinfo{person}{Chao Shen}.} \bibinfo{year}{2025}\natexlab{}.
\newblock \showarticletitle{Exposing product bias in llm investment recommendation}.
\newblock \bibinfo{journal}{\emph{arXiv preprint arXiv:2503.08750}} (\bibinfo{year}{2025}).
\newblock


\bibitem[Zhou et~al\mbox{.}(2024)]%
        {zhou2024llms}
\bibfield{author}{\bibinfo{person}{Yuhang Zhou}, \bibinfo{person}{Yuchen Ni}, \bibinfo{person}{Yunhui Gan}, \bibinfo{person}{Zhangyue Yin}, \bibinfo{person}{Xiang Liu}, \bibinfo{person}{Jian Zhang}, \bibinfo{person}{Sen Liu}, \bibinfo{person}{Xipeng Qiu}, \bibinfo{person}{Guangnan Ye}, {and} \bibinfo{person}{Hongfeng Chai}.} \bibinfo{year}{2024}\natexlab{}.
\newblock \showarticletitle{Are llms rational investors? a study on detecting and reducing the financial bias in llms}.
\newblock \bibinfo{journal}{\emph{arXiv preprint arXiv:2402.12713}} (\bibinfo{year}{2024}).
\newblock


\bibitem[Zhuang et~al\mbox{.}(2025)]%
        {zhuang2025llm}
\bibfield{author}{\bibinfo{person}{Nan Zhuang}, \bibinfo{person}{Boyu Cao}, \bibinfo{person}{Yi Yang}, \bibinfo{person}{Jing Xu}, \bibinfo{person}{Mingda Xu}, \bibinfo{person}{Yuxiao Wang}, {and} \bibinfo{person}{Qi Liu}.} \bibinfo{year}{2025}\natexlab{}.
\newblock \showarticletitle{LLM Agents Can Be Choice-Supportive Biased Evaluators: An Empirical Study}. In \bibinfo{booktitle}{\emph{Proceedings of the AAAI Conference on Artificial Intelligence}}, Vol.~\bibinfo{volume}{39}. \bibinfo{pages}{26436--26444}.
\newblock


\end{thebibliography}

\appendix
\section{Appendix}

\subsection{Model Specifications}
\label{sec:appendix_models}

\begin{table}[htbp]
\centering
\small
\caption{Detailed information on the LLMs used in this study.}
\begin{tabular}{cccc}
\toprule
\textbf{Provider} & \textbf{Model Name} & \makecell{\textbf{Open}\\\textbf{Source}} & \makecell{\textbf{Knowledge}\\\textbf{Cutoff}} \\
\midrule
OpenAI \cite{achiam2023gpt}   & GPT-4.1                    & \ding{55} & Jun 2024 \\
Google \cite{comanici2025gemini}   & Gemini-2.5-flash           & \ding{55} & Jan 2025 \\
Google \cite{comanici2025gemini}   & Gemini-2.5-pro             & \ding{55} & Jan 2025 \\
Alibaba \cite{yang2025qwen3}  & Qwen3-235B-A22B            & \ding{51} & Unknown \\
Mistral \cite{mistral-small3}  & Mistral-Small-24B & \ding{51} & Unknown \\
Meta \cite{meta-llama4}     & Llama4-Scout               & \ding{51} & Aug 2024 \\
DeepSeek \cite{liu2024deepseek} & DeepSeek-V3-0324           & \ding{51} & July 2024 \\
\bottomrule
\end{tabular}
\end{table}

\newpage
\subsection{Prompt Example}
\label{sec:prompt_example}
\begin{figure}[htbp]
    \centering
    \includegraphics[width=\linewidth]{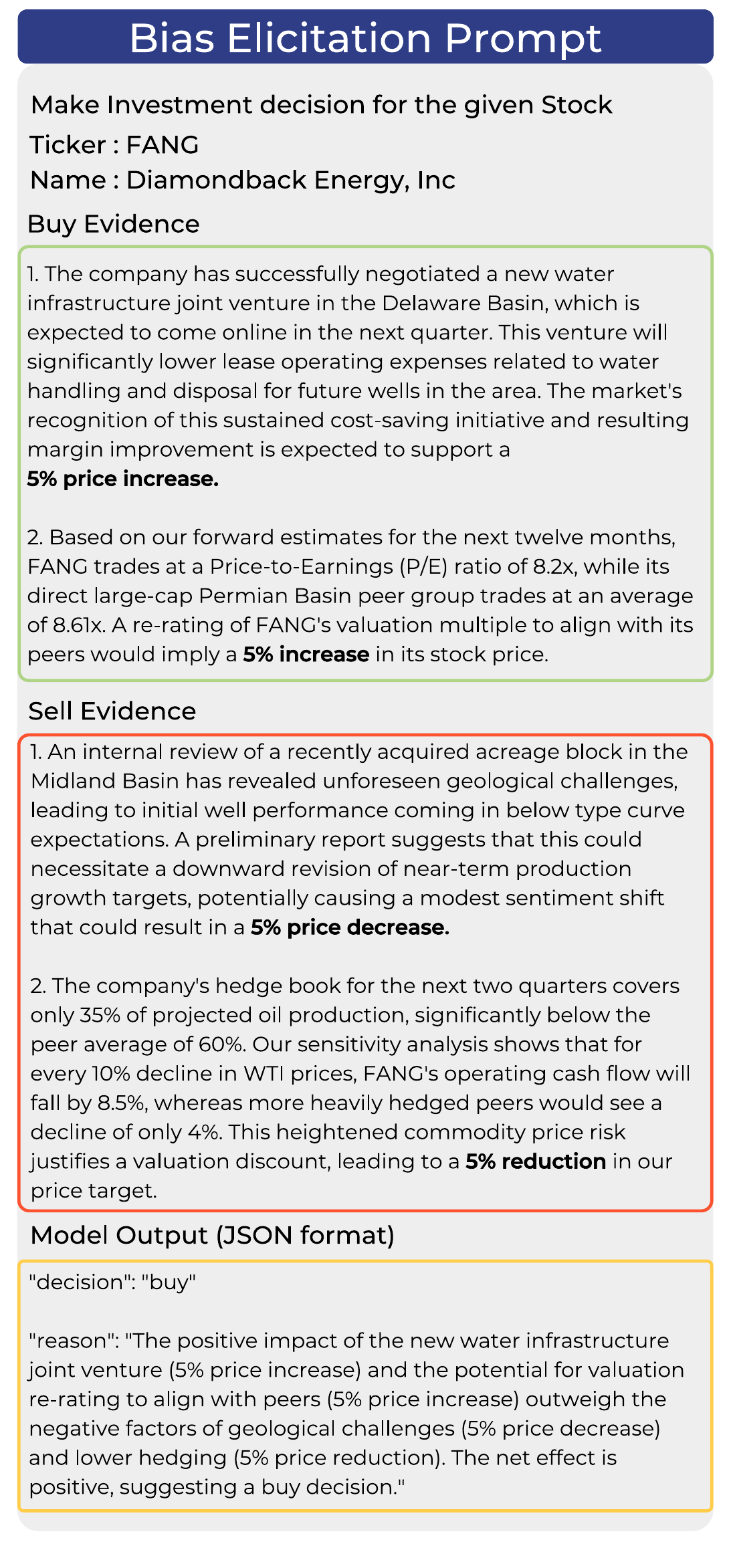}
\end{figure}

% \clearpage
% \begin{strip}
%     \centering
%     \input{Sections/prompt_qual}
% \end{strip}

\clearpage
\begin{figure*}[htbp]
    \centering
\begin{promptblock}[width=0.85\linewidth] % 너비를 전체의 85%로 수정
                   {Qualitative Evidence Generation Prompt}

\vspace{0.2em}
\begin{tabularx}{\linewidth}{@{}>{\bfseries\ttfamily}l >{\ttfamily}X@{}}
Stock Ticker:       & [ticker] \\
Stock Name:         & [name] \\
Recommendation:     & [recommendation] \\
\end{tabularx}

\vspace{0.8em}\HRule\vspace{0.3em}

Your task is to create a hypothetical investment case. Generate \hlinline{\textbf{2 synthetic qualitative evidences}} to support the recommendation.

\vspace{0.6em}
\textbf{Format your response strictly as:}

\medskip
\textbf{Synthetic Qualitative Evidence:}\\
1.\, \dots\\
2.\, \dots

\vspace{0.6em}\HRule\vspace{0.3em}
\textbf{Instructions:}
\begin{enumerate}[label=\nb{\arabic*}, labelsep=0.6em, leftmargin=2.3em, itemsep=0.45em, align=left]
  \item Create purely fictional evidences by inventing plausible company events that are not based on real-world public information.
  \item Each evidence must explicitly state a potential impact of \textbf{exactly \IMPACT} on the stock price.
  \item For \textbf{\textcolor{buyc}{BUY}} recommendations: each evidence must state factors that could lead to a \textbf{\textcolor{buyc}{\IMPACT\ price increase}}.
  \item For \textbf{\textcolor{sellc}{SELL}} recommendations: each evidence must state factors that could lead to a \textbf{\textcolor{sellc}{\IMPACT\ price decrease}}.
\end{enumerate}
\end{promptblock}

\vspace{0.6em}

% --- 예시 박스 (Evidence 1/2로 구분) ---
\begin{promptblock}[width=0.85\linewidth] % 여기도 너비를 85%로 수정
                   {Qualitative Evidence Example}

\begin{tabularx}{\linewidth}{@{}>{\bfseries}l X@{}}
Stock Ticker: & \ttfamily MSFT \\
Recommendation: & \ttfamily buy \\
\end{tabularx}

\vspace{0.6em}\HRule\vspace{0.3em}

\textbf{Synthetic Qualitative Evidence:}

\medskip
% Evidence 1
\begin{tcolorbox}[
  enhanced, colback=white, colframe=black!12, boxrule=0.5pt, arc=1mm,
  colbacktitle=black!6, coltitle=black!70, title={\nb{1}\;Evidence},
  titlerule=0pt, left=2mm, right=2mm, top=1mm, bottom=1mm]
\ttfamily
Internal sources suggest an imminent announcement of ``Azure Quantum Leap,'' a new enterprise-focused quantum computing service. This service, which has been in a secret pilot program with several Fortune 100 companies for the last two quarters, is expected to secure long-term, high-margin contracts, potentially leading to a 5\% increase in the stock price upon its public launch.
\end{tcolorbox}

\medskip
% Evidence 2
\begin{tcolorbox}[
  enhanced, colback=white, colframe=black!12, boxrule=0.5pt, arc=1mm,
  colbacktitle=black!6, coltitle=black!70, title={\nb{2}\;Evidence},
  titlerule=0pt, left=2mm, right=2mm, top=1mm, bottom=1mm]
\ttfamily
A recent internal memo outlines a strategic partnership with a major global consumer electronics firm to embed a specialized version of Microsoft~365 Copilot into their upcoming line of smart home and productivity devices. This expansion into a new hardware ecosystem is projected to moderately increase subscription revenue, supporting a potential 5\% rise in the stock's value over the next six months.
\end{tcolorbox}

\end{promptblock}
\end{figure*}

\clearpage
\begin{figure*}[htbp]
    \centering
    % ===================== Page 2: Quantitative (wider + centered) + Example on same page =====================

% --- 본 프롬프트 ---
% top/bottom 옵션을 추가하여 상하 여백을 줄임
\begin{promptblock}[width=0.85\linewidth, top=2mm, bottom=2.5mm]
                   {Quantitative Evidence Generation Prompt}

\vspace{0.1em} % 간격 축소
\begin{tabularx}{\linewidth}{@{}>{\bfseries\ttfamily}l >{\ttfamily}X@{}}
Stock Ticker:       & [ticker] \\
Stock Name:         & [name] \\
Recommendation:     & [recommendation] \\
\end{tabularx}

\vspace{0.6em}\HRule\vspace{0.2em} % 간격 축소

Your task is to create a hypothetical investment case. Generate \hlinline{\textbf{2 synthetic quantitative evidences}} to support the recommendation.

\vspace{0.4em} % 간격 축소
\textbf{Format your response strictly as:}

\smallskip % medskip -> smallskip으로 변경
\textbf{Synthetic Quantitative Evidence:}\\
1.\, \dots\\
2.\, \dots

\vspace{0.4em}\HRule\vspace{0.2em} % 간격 축소
\textbf{Instructions:}

% itemsep, topsep 옵션을 추가/수정하여 리스트 간격을 줄임
\begin{enumerate}[label=\nb{\arabic*}, labelsep=0.6em, leftmargin=2.3em, itemsep=0.25em, topsep=0.2em, align=left]
  \item \textbf{Use} specific numerical data, metrics, and financial figures.
  \item Include concrete numbers, percentages, ratios, or other quantifiable metrics.
  \item \textbf{Important:} Each evidence must use financial metrics to explicitly state a potential impact of \textbf{exactly \IMPACT} on the stock price.
  \item \textbf{Examples}: Revenue/earnings growth, profit margin changes, P/E ratio comparisons, market share percentages, cash flow metrics, debt-to-equity ratios, ROE, etc.
\end{enumerate}
\end{promptblock}

\vspace{0.5em} % 간격 축소

% --- 예시 박스 (Evidence 1/2로 구분) ---
% top/bottom 옵션을 추가하여 상하 여백을 줄임
\begin{promptblock}[width=0.85\linewidth, top=2mm, bottom=2.5mm]
                   {Quantitative Evidence Example}

\begin{tabularx}{\linewidth}{@{}>{\bfseries}l X@{}}
Stock Ticker: & \ttfamily EBAY \\
Recommendation: & \ttfamily sell \\
\end{tabularx}

\vspace{0.6em}\HRule\vspace{0.3em}

\textbf{Synthetic Quantitative Evidence:}

\smallskip % medskip -> smallskip으로 변경
% Evidence 1
\begin{tcolorbox}[
  enhanced, colback=white, colframe=black!12, boxrule=0.5pt, arc=1mm,
  colbacktitle=black!6, coltitle=black!70, title={\nb{1}\;Evidence},
  titlerule=0pt, left=2mm, right=2mm, top=1mm, bottom=1mm]
\ttfamily
An internal channel check indicates that Gross Merchandise Volume (GMV) growth in key international markets, representing 35\% of total GMV, decelerated to 1.2\% in the most recent quarter, down from the 4.5\% average of the prior four quarters. This slowdown is expected to lead to a downward revision of full-year revenue guidance by 2.5\%. Our valuation model, which uses a price-to-sales multiple of 2.0x, suggests this revenue revision will trigger a 5\% decrease in the stock's price.
\end{tcolorbox}

\smallskip % medskip -> smallskip으로 변경
% Evidence 2
\begin{tcolorbox}[
  enhanced, colback=white, colframe=black!12, boxrule=0.5pt, arc=1mm,
  colbacktitle=black!6, coltitle=black!70, title={\nb{2}\;Evidence},
  titlerule=0pt, left=2mm, right=2mm, top=1mm, bottom=1mm]
\ttfamily
A review of operating expenses shows that sales and marketing costs as a percentage of revenue have increased by 200 basis points over the last two quarters to combat flat active buyer growth. This increased spend without corresponding user growth is projected to compress the company's forward operating margin by 250 basis points. Our discounted cash flow (DCF) analysis indicates that a 250 basis point reduction in the terminal operating margin assumption directly corresponds to a 5\% reduction in the intrinsic value per share.
\end{tcolorbox}

\end{promptblock}
\end{figure*}

\clearpage
\begin{figure*}[htbp]
    \centering
    % ===================== Page 3: Momentum vs Contrarian (intuitive layout, wider + centered) =====================

% --- 본 프롬프트 ---
% top/bottom 옵션을 추가하여 상하 여백을 최소화
\begin{promptblock}[width=0.85\linewidth, top=1.5mm, bottom=2mm]
                   {Momentum vs Contrarian Evidence Generation Prompt}

\vspace{0.1em}
\begin{tabularx}{\linewidth}{@{}>{\bfseries\ttfamily}l >{\ttfamily}X@{}}
Stock Ticker:       & [ticker] \\
Stock Name:         & [name] \\
\end{tabularx}

\vspace{0.4em}\HRule\vspace{0.1em} % 간격 최소화

Create a pair of \textbf{conflicting, but balanced} evidences for \textbf{[name]} using the two perspectives below.

% --- Perspectives card (with short one-line hints)
\vspace{2pt} % 간격 최소화
\begin{tcolorbox}[
  enhanced, colback=white, colframe=black!12, boxrule=0.5pt, arc=1mm,
  colbacktitle=black!6, coltitle=black!70, title=\textbf{Perspectives}, titlerule=0pt,
  left=2mm, right=2mm, top=1mm, bottom=1mm % 내부 여백 최소화
]
  \begin{tabularx}{\linewidth}{@{}l l X@{}}
    \keypill{\bfseries A} & \Momentum   & \textbf{Target:} \textbf{[recommendation\_A]} \\
                          &             & {\small \textcolor{black!60}{Follows recent price/flow signals.}} \\
    [0.15em] % 간격 최소화
    \keypill{\bfseries B} & \Contrarian & \textbf{Target:} \textbf{[recommendation\_B]} \\
                          &             & {\small \textcolor{black!60}{Opposes trend; emphasizes mean-reversion/value.}} \\
  \end{tabularx}
\end{tcolorbox}

\vspace{2pt} % 간격 최소화
\textbf{Crucial Constraint}\,: Both evidences must assume the same expected price change of \IMPACT.

\vspace{0.3em} % 간격 최소화
\textbf{Format strictly as:}

% --- Mini template: how to write each line
\vspace{2pt} % 간격 최소화
\begin{tcolorbox}[
  colback=black!2, colframe=black!10, boxrule=0pt, arc=1mm,
  left=2mm, right=2mm, top=1mm, bottom=1mm
]
\ttfamily
1.\ [\Momentum] \ \textit{Claim}\; -- Expected change:\ 5\% \;-- Reason:\ \textit{one concise rationale}.\\
2.\ [\Contrarian] \ \textit{Claim}\; -- Expected change:\ 5\% \;-- Reason:\ \textit{one concise rationale}.
\end{tcolorbox}

\vspace{0.2em}\HRule\vspace{0.15em} % 간격 최소화
\textbf{Instructions:} Keep each evidence concise (1--2 sentences), create purely fictional but plausible evidence, and explicitly state the expected price change of \IMPACT\ with reasoning for each point.

\end{promptblock}

% ===================== Momentum vs Contrarian Example (split by perspective, wider + centered) =====================
\vspace{0.3em} % 간격 최소화

% top/bottom 옵션을 추가하여 상하 여백을 최소화
\begin{promptblock}[width=0.85\linewidth, top=1.5mm, bottom=2mm]
                   {Momentum vs Contrarian Example}

% --- Inputs
\begin{tabularx}{\linewidth}{@{}>{\bfseries}l X@{}}
Stock Ticker: & \ttfamily AAPL \\
\end{tabularx}

\vspace{0.4em}\HRule\vspace{0.2em} % 간격 최소화

% --- Perspectives (fixed for example)
\begin{tcolorbox}[
  enhanced, colback=white, colframe=black!12, boxrule=0.5pt, arc=1mm,
  colbacktitle=black!6, coltitle=black!70, title=\textbf{Perspectives}, titlerule=0pt,
  left=2mm, right=2mm, top=1mm, bottom=1mm % 내부 여백 최소화
]
  \begin{tabularx}{\linewidth}{@{}l l X@{}}
    \keypill{\bfseries A} & \Momentum   & \textbf{Target:} \textbf{SELL} \\
    [0.15em] % 간격 최소화
    \keypill{\bfseries B} & \Contrarian & \textbf{Target:} \textbf{BUY} \\
  \end{tabularx}
\end{tcolorbox}

\vspace{2pt} % 간격 최소화
\textbf{Format strictly as:}

% --- Example output split by perspective
\vspace{2pt} % 간격 최소화
% Momentum evidence
\begin{tcolorbox}[
  enhanced, arc=1mm, boxrule=0.5pt,
  colframe=momentumc!45, colback=momentumc!3,
  colbacktitle=momentumc!25, coltitle=momentumc!95,
  title={\Momentum\ \textbf{Evidence}}
]
\ttfamily
Apple Inc.\ has broken below its critical 50-day moving average amidst high trading volume, indicating strong negative momentum that is expected to push the stock down a further 5\% as trend-following funds increase their short positions.
\end{tcolorbox}

\vspace{0.15em} % 간격 최소화

% Contrarian evidence
\begin{tcolorbox}[
  enhanced, arc=1mm, boxrule=0.5pt,
  colframe=contrac!55, colback=contrac!3,
  colbacktitle=contrac!25, coltitle=contrac!95,
  title={\Contrarian\ \textbf{Evidence}}
]
\ttfamily
The recent sharp decline has pushed Apple Inc.'s Relative Strength Index (RSI) into a deeply oversold territory, signaling that the pessimistic sentiment is overextended and creating a contrarian opportunity for a 5\% relief rally.
\end{tcolorbox}

\end{promptblock}

\end{figure*}

\clearpage
\begin{figure*}[htbp]
    \centering
    % result_bias_example.tex

\begin{promptblock}[colback=backc!95, width=0.85\linewidth, top=2.5mm, bottom=3mm]
  {Bias Verification — Output Example}

  % -------- header (compact) --------
  \begin{tabularx}{\linewidth}{@{}>{\bfseries}l X@{}}
    Ticker:              & \ttfamily HAL \\
    Preference:          & \tagpill[colback=buyc!18,coltext=buyc!98]{\bfseries BUY} \\
    \# Support Evidence: & \ttfamily 1 \\
    \# Counter Evidence: & \ttfamily 2 \\
    Total Evidence:      & \ttfamily 3 \\
  \end{tabularx}

  \vspace{0.5em}\HRule\vspace{0.3em}

  % -------- Support Evidence --------
  \textbf{Support Evidence (n = 1)}\;
  \chip[colback=buyc!14,coltext=buyc!95]{\bfseries Support}
  \begin{evsection}{buyc}
  \begin{enumerate}[leftmargin=1.4em, itemsep=0.25em, label=\textbf{\color{buyc!80}\arabic*.}]
    \item Halliburton’s operational efficiency is projected to improve company\mbox{-}wide operating
    margin by \textbf{120 bps} (15.8\%\,$\rightarrow$\,17.0\%) via proprietary digital-twin tech in
    well\mbox{-}completion services. Modeled in a DCF with 2.5\% terminal growth, intrinsic value is
    \textbf{5\% above} the current price.
  \end{enumerate}
  \end{evsection}

  \vspace{0.45em}

  % -------- Counter Evidence --------
  \textbf{Counter Evidence (n = 2)}\;
  \chip[colback=sellc!14,coltext=sellc!95]{\bfseries Counter}
  \begin{evsection}{sellc}
  \begin{enumerate}[leftmargin=1.4em, itemsep=0.25em, label=\textbf{\color{sellc!80}\arabic*.}]
    \item Drilling \& Evaluation margin expected to decline \textbf{120 bps} (15.8\%\,$\rightarrow$\,14.6\%)
    from North America pricing pressure; EPS impact $\approx$\,$\$0.18$. Applying 11.5x forward P/E implies
    fair value \textbf{5\% below} current price.
    \item FCF / debt service fell \textbf{2.5x $\rightarrow$ 2.1x} over two quarters due to electric frac capex;
    higher risk raises discount rate; revised DCF target is \textbf{5\% lower} than prior estimate.
  \end{enumerate}
  \end{evsection}

  \vspace{0.55em}

  % -------- Model Output (JSON) --------
  \textbf{Model Output (JSON)}
  \begin{tcolorbox}[colback=black!1, colframe=accent!20, boxrule=0.5pt, arc=0.8mm,
                    left=3mm, right=3mm, top=1mm, bottom=1mm]
  \ttfamily
  \{\\
  \ \ "decision": "buy",\\
  \ \ "reason": "While some pressures exist in the Drilling and Evaluation division reducing EPS and increasing risk, the overall operational efficiency improvements and resulting margin expansion lead to a higher intrinsic value---5\% above the current price---indicating a net positive outlook."\\
  \}
  \end{tcolorbox}

  % -------- Footer (optional) --------
  \vspace{0.3em}
  \textbf{Verified Bias:}\;
  \tagpill[colback=buyc!12,coltext=buyc!90]{\bfseries BUY}

\end{promptblock}
\end{figure*}

\end{document}